\documentclass[sn-nature]{sn-jnl}

\usepackage{graphicx}%
\usepackage{multirow}%
\usepackage{amsmath,amssymb,amsfonts}%
\usepackage{amsthm}%
\usepackage{mathrsfs}%
\usepackage[title]{appendix}%
\usepackage{xcolor}%
\usepackage{textcomp}%
\usepackage{manyfoot}%
\usepackage{booktabs}%
\usepackage{algorithm}%
\usepackage{algorithmicx}%
\usepackage{algpseudocode}%
\usepackage{listings}%
\usepackage{siunitx}
\usepackage{svg}

\theoremstyle{thmstyleone}%
\theoremstyle{thmstyletwo}%

\theoremstyle{thmstylethree}%

\begin{document}

\title[Article Title]{\textbf{Scalable, nanoscale positioning of highly coherent color centers in prefabricated diamond nanostructures}}

\author[1]{\fnm{Sunghoon} \sur{Kim}}
\author[1]{\fnm{Paz} \sur{London}}
\author[1]{\fnm{Daipeng} \sur{Yang}}
\author[1]{\fnm{Lillian B.} \sur{Hughes}}
\author[1]{\fnm{Jeffrey} \sur{Ahlers}}
\author[1]{\fnm{Simon} \sur{Meynell}}
\author[2]{\fnm{William J.} \sur{Mitchell}}
\author[3]{\fnm{Kunal} \sur{Mukherjee}}
\author*[1]{\fnm{Ania C. Bleszynski} \sur{Jayich}}\email{ania@physics.ucsb.edu}

\affil[1]{Department of Physics, University of California, Santa Barbara, Santa Barbara, California 93106, USA}

\affil[2]{Nanofabrication Facility, Department of Electrical and Computer Engineering, University of California, Santa Barbara, Santa Barbara, California 93106, USA}

\affil[3]{Department of Materials Science and Engineering, Stanford University, Palo Alto, CA 94305, USA}

\abstract{
Nanophotonic devices in color center-containing hosts provide efficient readout, control, and entanglement of the embedded emitters. 
Yet control over color center formation -- in number, position, and coherence -- in nanophotonic devices remains a challenge to scalability. 
Here, we report a controlled creation of highly coherent diamond nitrogen-vacancy (NV) centers with nanoscale three-dimensional localization in prefabricated nanostructures with high yield.
Combining nitrogen $\delta$-doping during chemical vapor deposition diamond growth and localized electron irradiation, we form shallow NVs registered to the center of diamond nanopillars with wide tunability over NV number. 
We report positioning precision of $\sim$ \qty{4}{\nano\metre} in depth and \qty{46(1)}{\nano\meter} laterally in pillars (\qty{102(2)}{\nano\meter} in bulk diamond). 
We reliably form single NV centers with long spin coherence times (average $T_{2}^{Hahn}$ = \qty{98}{\micro\second}) and 1.8\(\times\) higher average photoluminescence compared to NV centers randomly positioned in pillars. 
We achieve a 3\(\times\) improved yield of NV centers with single electron-spin sensitivity over conventional implantation-based methods. 
Our high-yield defect creation method will enable scalable production of solid-state defect sensors and processors.
}
\maketitle

\section*{Main}
Optically addressable solid-state spin defects are versatile tools for quantum-enhanced technologies\cite{awschalom_quantum_2018,wolfowicz_quantum_2021}. 
The photonic degree of freedom enables single-spin readout\cite{doherty_nitrogen-vacancy_2013} and control\cite{yale_all-optical_2013,hilser_all-optical_2012} and entanglement generation\cite{bernien_heralded_2013, stolk_metropolitan-scale_2024}.
Moreover, engineered nanophotonic structures can greatly enhance spin-photon interfaces, where customized structures 
 such as cavities\cite{tomljenovic-hanic_diamond_2006,lukin_4h-silicon-carbide--insulator_2020,
 babin_fabrication_2022, bhaskar_experimental_2020, stas_robust_2022,  wang_fabrication_2007}, solid immersion lenses\cite{marseglia_nanofabricated_2011}, metalenses\cite{huang_monolithic_2019}, nanobeams \cite{khanaliloo_single-crystal_2015,babin_fabrication_2022} or nanowires \cite{   zhou_scanning_2017, babinec_diamond_2010} can be fabricated in the host material to increase collection efficiency\cite{jamali_microscopic_2014}, waveguide emitted photons\cite{lukin_4h-silicon-carbide--insulator_2020} or Purcell-enhance photon emission\cite{gadalla_enhanced_2021, crook_purcell_2020,bracher_selective_2017,faraon_coupling_2012, hausmann_coupling_2013,riedrich-moller_nanoimplantation_2015}.
In particular, diamond color centers are readily interfaced with engineered photonic structures to provide these advanced functionalities 
\cite{zhang_high-fidelity_2021,evans_photon-mediated_2018,pingault_all-optical_2014, becker_ultrafast_2016}. 
To realize efficient defect-photon interfaces, it is necessary to engineer a good spatial overlap between the optical mode of the nanostructured device and the defect. 
However, control over the formation of color centers in position and number, while maintaining reproducibly long spin coherence, remains an outstanding problem in realizing scalable fabrication of devices equipped with quantum-enhanced functionalities. 

Conventionally, color centers are created via ion implantation prior to device fabrication. 
The implantation dosage is chosen to match a target number of defects per device, but placement is random in the nanostructures. Subsequently, in a time-intensive, low-yield, and hence nonscalable post-selection process, devices with defects at the ideal position (\textit{e.g.}, at the mode maximum of the optical field) are selected. 
Moreover, the properties of the selected defects can degrade during the subsequent device fabrication process, \textit{e.g.}, at the etching step\cite{cui_reduced_2015}. 
Alternate approaches utilize highly specialized, home-built localized implantation techniques to spatially co-locate a defect and a nanostructure.  Atomic force microscopy-assisted implantation\cite{lesik_maskless_2013,pezzagna_nanoscale_2010} and focused ion beam implantation\cite{schroder_scalable_2017} have demonstrated lateral confinement inside prefabricated nanostructures to $\sim$ \(\qty{20}-\qty{30}{\nano \meter}\). Recently, a patterning technique \cite{wang_self-aligned_2022} involving implantation masks has shown \(\sim \qty{15}{\nano\meter}\) lateral positioning precision in a nanopillar, though the technique is limited to nanopillar geometries. 

However, implantation-based techniques have several drawbacks. Most critically, they suffer from collateral damage incurred during implantation (\textit{e.g.}, vacancy clusters) that adversely affect optical\cite{van_dam_optical_2019, kasperczyk_statistically_2020} and spin properties\cite{momenzadeh_nanoengineered_2015,tetienne_spin_2018,wang_self-aligned_2022} of nitrogen-vacancy (NV) centers. For instance, spin coherence times have been limited to \(T_2^{Hahn}\) less than 20 and \qty{50}{\micro \second} for \(> \qty{90}{\percent}\) of 10 and \( \qty{15}{\kilo e \volt}\) implanted NV centers, respectively\cite{momenzadeh_nanoengineered_2015,wang_self-aligned_2022}. Damage is also exacerbated at higher implantation dosages\cite{tetienne_spin_2018}, which are necessary for, \textit{e.g.}, achieving high defect densities or ensuring the presence of a defect in a small target volume. Further, the spread in the depth of implanted defects hinders their precise vertical positioning. For example, \( \qty{15}{\kilo e \volt}\) implanted NV centers have a vertical spread of \(>\)\qty{14}{\nano\meter} due to implantation straggling\cite{pezzagna_creation_2010} and ion channeling\cite{toyli_chip-scale_2010}, effects that become even more severe at higher implantation energies.

In contrast to ion implantation, nitrogen $\delta$-doping during chemical vapor deposition (CVD) diamond growth enables NV formation with reproducibly long spin coherence, nanometer-scale depth confinement even at large depths, and independent tunability over a wide range of nitrogen and NV densities\cite{hughes_two-dimensional_2023,hughes_strongly_2024,ohno_engineering_2012,ohno_three-dimensional_2014, eichhorn_optimizing_2019}. 
Previous studies on $\delta$-doped CVD-grown diamond demonstrated NV center densities tunable from $<$1 to \qty{47}{\unit
{ppm \cdot \nano\meter}} 
 using electron irradiation\cite{mclellan_patterned_2016, hughes_two-dimensional_2023,hughes_strongly_2024}, $<$\(\qty{3}{\nano \meter}\) depth confinement, and reproducibly long coherence times, with even \qty{15}{\nano \meter}-deep NVs showing \(T_2^{Hahn} > \) \qty{100}{\micro \second}\cite{myers_probing_2014}. Depth confinement of NV centers using $\delta$-doping has been used to enhance their coupling to nanophotonic devices, such as photonic crystal nanobeam cavities\cite{lee_deterministic_2014}, but without controlled lateral positioning.
Local vacancy creation techniques \cite{chen_laser_2017,kurita_efficient_2018,chen_laser_2019,stephen_deep_2019,fujiwara_creation_2023,shimotsuma_formation_2023} can also provide lateral confinement in addition to the depth confinement afforded by $\delta$-doping, but this capability has only been demonstrated on a bulk substrate without alignment to prefabricated photonic structures \cite{ohno_three-dimensional_2014,mclellan_patterned_2016}.

In this paper, we demonstrate high-throughput, localized formation of highly coherent NV centers aligned to prefabricated nanophotonic structures. 
We register a \qty{200}{\kilo\electronvolt} electron beam with \qty{20}{\nano\meter} spot size, which we call $\delta$-electron irradiation, to the center of diamond nanopillars fabricated in CVD-grown diamond with a \qty{53}{\nano\meter}-deep, $\delta$-doped nitrogen layer. By controlling the electron dose and annealing time, we tune the average number of NVs per irradiation spot inside nanopillars from \(\ \sim 0\) to 10. We report lateral confinement of created NVs to a standard deviation of \( \qty{102(2)}{\nano \meter}\) in unpatterned diamond in addition to \(\sim\) \qty{4}{\nano \meter} vertical confinement.
Lateral confinement is improved to \qty{46(1)}{\nano \meter} in 280 nm diameter pillars and \qty{72(1)}{\nano \meter} in 480 nm diameter pillars.
We find that our observations agree well with Monte Carlo (MC) simulations based on a simple diffusion-capture model.
Importantly, the NVs formed using our method feature repeatably long spin coherence time (average $T_{2}^{Hahn}$ = \qty{98 (37)}{\micro\second}) with a high spin-dependent photoluminescence ($PL$) contrast of \qty{18 (4)}{\percent}. Additionally, we observe a 1.8\(\times\) enhancement of $PL$ from NV centers localized to pillars compared to pillars with non-localized NVs.
Lastly, we demonstrate \(\times 3\) increase in the expected yield of single electron spin-sensitive magnetometers compared to conventional methods. Overall, this technique facilitates the scalable fabrication of state-of-the-art solid-state defect-assisted devices, where scalable refers to a high-yield, time-efficient process that leverages commercially available tools.

\subsection*{Targeted formation of NVs in nanostructures}
\begin{figure}
    \centering
    \centerline{\includegraphics[width= 89mm]{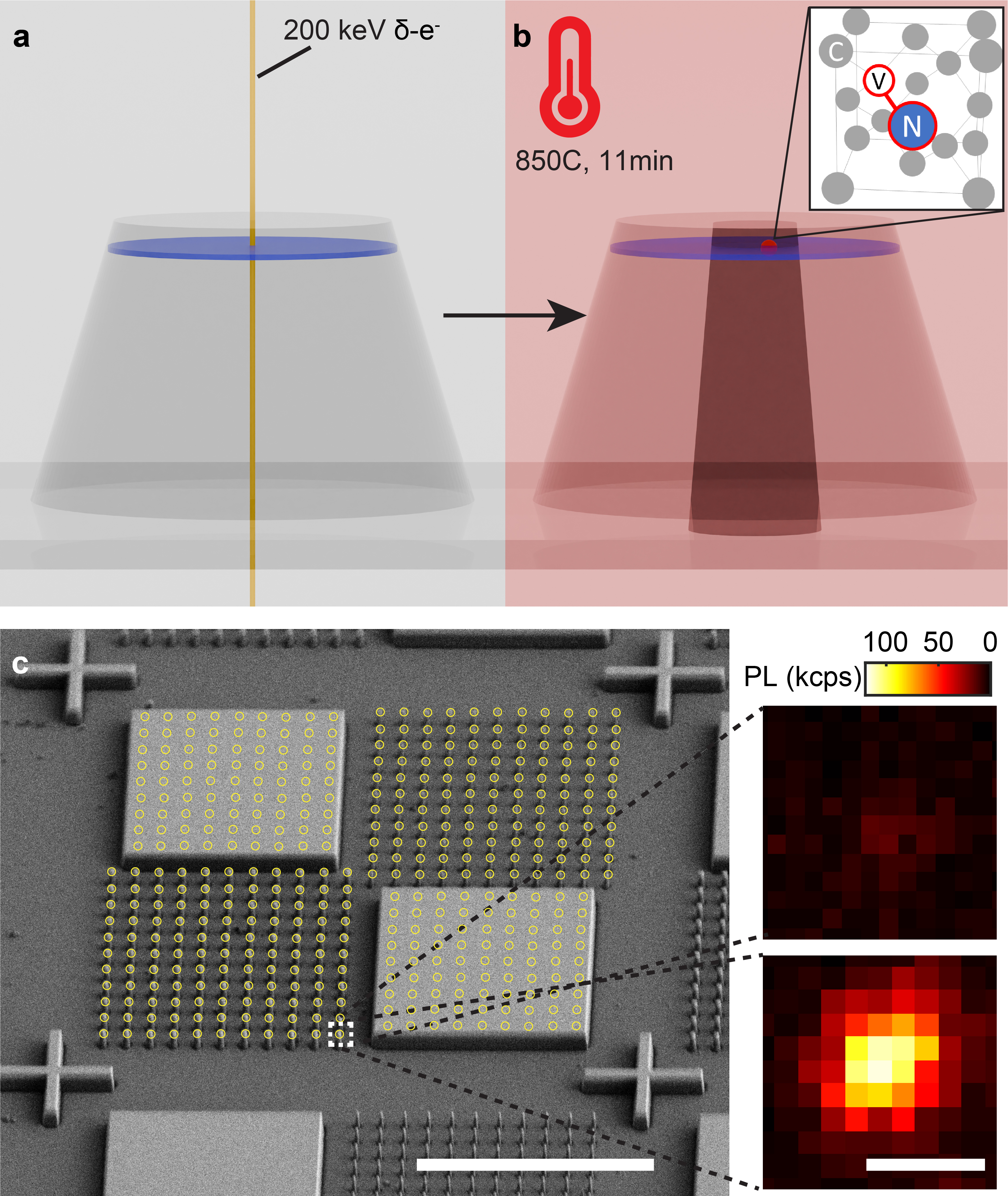}}
    \caption{Targeted formation of color centers aligned to prefabricated diamond nanostructures. (a) Schematic of our formation method showing a nanopillar 
    containing a near-surface nitrogen $\delta$-doped layer (blue disk). In $\delta$-electron irradiation, an electron beam of \qty{20}{\nano\meter} spot size (yellow line) irradiates the center of a pillar, creating vacancies along its trajectory. (b) Upon annealing, monovacancies diffuse to form a vacancy-rich region (dark shaded region) in which they can be captured by nitrogen atoms to form NV centers (inset). 
    (c) A scanning electron micrograph of a unit block of etched diamond pillars and mesas framed by alignment marks. Each $50\times$\qty{ 50}{\micro \meter\squared} unit block consists of two unetched mesas (top left and bottom right), and two regions of nanopillars with a diameter of \(\qty{280}{\nano \meter}\) (top right) and \(\qty{480}{\nano \meter}\)  (bottom left). The overlaid yellow circles denote the target positions of the electron beam (scale bar: \qty{20}{\micro \meter}). The insets show confocal $PL$ images of a single \(\qty{480}{\nano \meter}\) pillar before (top) and after (bottom) \qty{4.8e21}{\elementarycharge^-\per\centi\meter\squared} $\delta$-electron irradiation and annealing (scale bar: \qty{1}{\micro \meter}). }
    \label{fig1}
\end{figure}

Our targeted formation of NVs in prefabricated nanostructure utilizes localized electron irradiation and timed vacuum annealing, as shown schematically in Figure~\ref{fig1}(a-b).
 We first fabricate our device from a CVD-grown diamond with a \qty{53}{\nano \meter}-deep $^{15}$N $\delta$-doped layer, as described in Methods \cite{hughes_two-dimensional_2023}. We use electron beam lithography (EBL) followed by inductively coupled plasma-reactive ion etching (ICP-RIE) recipes to transfer nanostructures onto the diamond substrate. We use a negative electron beam resist (FOx-16, Dow Corning) as a mask for etching \(\approx \qty{1}{\micro \meter}\) tall features with Ar/O$_2$ plasma. Nanopillars with diameters \qty{280}{\nano\meter} and \qty{480}{\nano\meter}, square mesas with \qty{20}{\micro\meter} \(\times\) \qty{20}{\micro\meter} dimensions, and alignment marks were fabricated, as shown in the scanning electron micrograph (SEM) in Figure~\ref{fig1}(c). 
 
To laterally localize NV centers, we use a commercially available \qty{200}{\kilo \electronvolt} EBL tool (JBX-8100FS, JEOL Ltd.) to $\delta$-electron irradiate the centers of the nanopillars using an electron beam of \qty{20}{\nano \meter} spot size (Figure~\ref{fig1}(a)); these electrons can penetrate into the diamond and displace carbon atoms along their trajectory up to \(\sim \qty{50}{\micro\meter}\) below the surface\cite{campbell_radiation_2000, mclellan_patterned_2016}, creating a narrow pencil of vacancies(Figure~\ref{fig1}(a), SI). 
We note that \qty{145}{\kilo \electronvolt} is the threshold energy for vacancy creation in diamond \cite{mclellan_patterned_2016} and only recently have commercial EBL tools exceeded \qty{150}{\kilo\volt}. 
The resulting monovacancy density depends on electron dose, which we tune from \qty{1.6e19}{\elementarycharge^-\per\centi\meter\squared} to \qty{4.8e21}{\elementarycharge^-\per\centi\meter\squared} by adjusting the dwell time while keeping the beam current constant at \qty{20}{\nano\ampere}. 
Subsequent annealing at \qty{850}{\celsius} for 11 minutes in vacuum promotes the diffusion of monovacancies (Figure~\ref{fig1}(b)). When a monovacancy diffuses to a site adjacent to a nitrogen atom, it can get captured to form an NV center (Figure~\ref{fig1}(b), inset). We note that NVs formed in the $\delta$-doped N-layer can be identified by their \(^{15}\)NV hyperfine structure, revealed with pulsed electron spin resonance (ESR) spectroscopy. Then, the device is cleaned in a boiling nitrating acid (1:1 HNO$_3$:H$_2$SO$_4$) and annealed at \qty{450}{\celsius} in air.
All NV measurements were taken using a home-built confocal microscope with \qty{532}{\nano \meter} excitation\cite{bluvstein_identifying_2019}.

\begin{figure}
    
    \centering
    \includegraphics[width= 88mm]{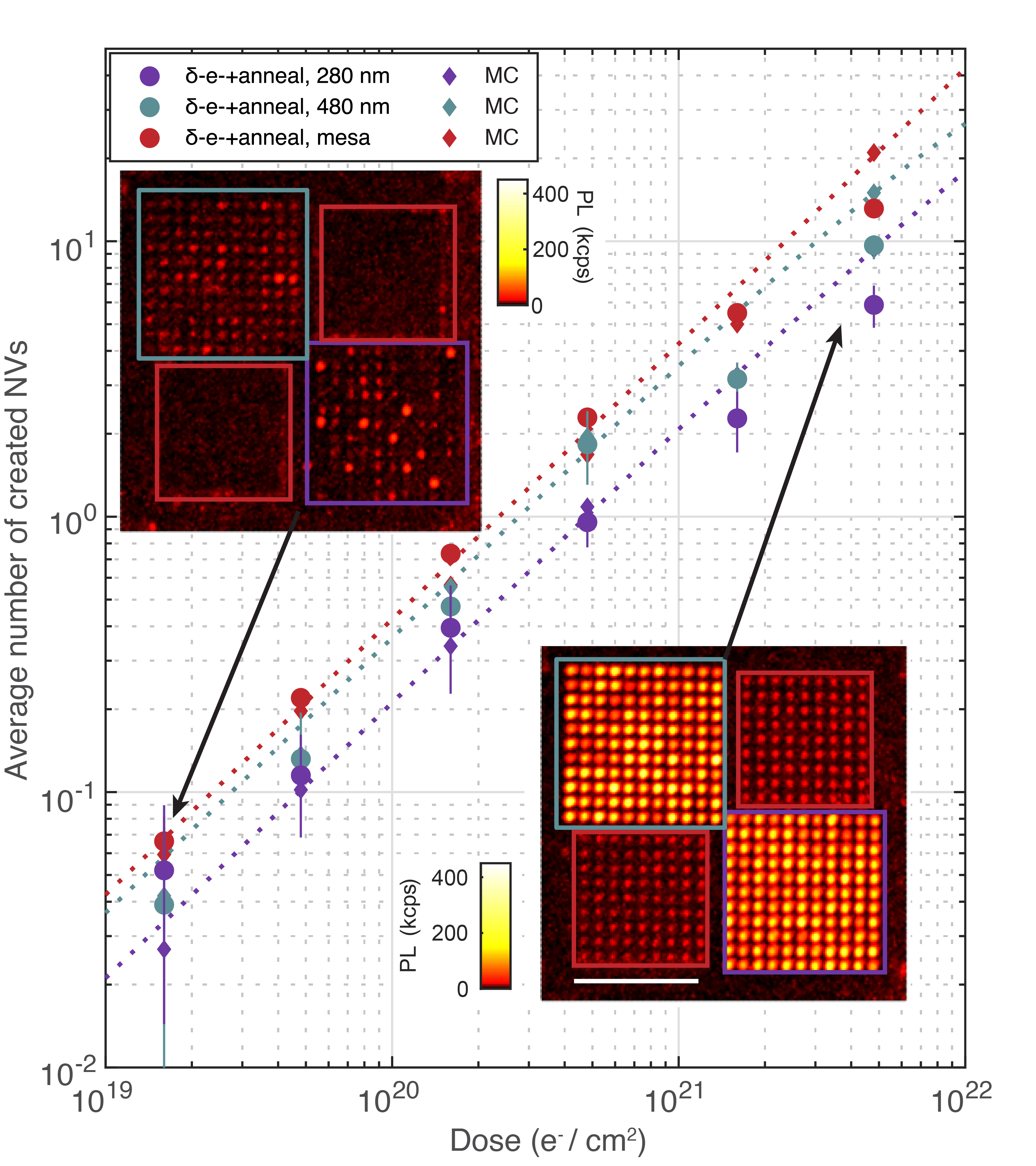}
    \caption{Electron dose control over NV creation. Plotted are the average number of created NVs per spot in  
    \qty{280}{\nano \meter} diameter pillars (purple circles), in \qty{480}{\nano \meter} diameter pillars (teal circles), and in the mesas (red circles). The error bars denote \qty{95}{\percent} confidence interval of the NV number estimation.
    The results from MC simulations (diamonds) are plotted in corresponding colors and show good agreement with the measurements. The fitted curves for the MC simulation results (dotted lines) are shown as a guide to the eye. The insets show confocal micrographs of a unit block after irradiation and annealing with a dose of  \qty{1.6e19}{\elementarycharge^-\per\centi\meter\squared} (top left) and \qty{4.8e21}{\elementarycharge^-\per\centi\meter\squared} (bottom right). (scale bar: \qty{20}{\micro \meter}). The overlaid boxes indicate the locations of the \qty{280}{\nano \meter} pillars (purple box), \qty{480}{\nano \meter} pillars (teal box), and mesas (red boxes). }
    \label{fig2}
\end{figure}

First, we show control over the number of NVs formed per nanopillar by varying the $\delta$-electron irradiation dosage. In Figure~\ref{fig2}, we sweep the irradiation dose from \qty{1.6e19}{\elementarycharge^-\per\centi\meter\squared} to \qty{4.8e21}{\elementarycharge^-\per\centi\meter\squared} and measure the average number of NVs created in \qty{280}{\nano \meter} (purple circles) and \qty{480}{\nano \meter} (teal circles) diameter pillars. We use a maximum likelihood estimation (MLE) method \cite{mclellan_patterned_2016} based on continuous wave-ESR spectroscopy (CW-ESR) taken on 121 pillars for each pillar size and irradiation dose after annealing (details in SI). With increasing irradiation dose, the average number of created NVs increases monotonically from \(\ \sim 5\times 10^{-2}\) to 5.9(7) and 9.7(4) for \qty{280}{\nano\meter} and \qty{480}{\nano \meter} diameter pillars, respectively. We subtract the contribution from as-grown NVs, characterized by measuring non-irradiated pillars of each diameter (see SI). 
We also estimate the average NV number per spot in $\delta $-electron irradiated mesas by measuring the total $PL$ around the target areas normalized by that of single NVs, as plotted in red circles in Figure~\ref{fig2}. Likewise, we observe a monotonic increase in the NV formation with irradiation dose.

The increase in NV number with pillar diameter indicates that vacancies diffuse at least as far as the radius of the smaller pillar, and comparing our results to Monte Carlo (MC) simulations (Methods) of a simple diffusion-capture model of NV formation, we extract a monovacancy diffusion constant \(D_V =\) \qty{17(4)}{\nano\meter\squared \per \second} (see SI). Simulation results are plotted as diamonds in Figure~\ref{fig2}. At higher irradiation doses, (\(\geq \qty{1.6e21}{\elementarycharge^-\per\centi\meter\squared}\)) simulations slightly overestimate NV number, which we attribute to the creation of vacancy clusters at high monovacancy density\cite{iakoubovskii_vacancy_2004,davies_vacancy-related_1992,santonocito_nv_2024}, giving rise to sublinear monovacancy creation efficiency. 
In the next section, we characterize \(D_V\) via an alternate approach, arriving at a similar value, and we further discuss the results.

We identify $\sim$ \qty{5e20}{\elementarycharge^-\per\centi\meter\squared}and \qty{3e20}{\elementarycharge^-\per\centi\meter\squared} as the target dose for an average of 1 NV per \qty{280}{\nano\meter} and \qty{480}{\nano\meter} pillar, respectively, facilitating the fabrication of devices based on single isolated defects for sensing and networking applications. We expect the target electron dose to vary for different nitrogen densities and nanostructure geometries.

\begin{figure}
    \centering
    \centerline{\includegraphics[width= 180mm]{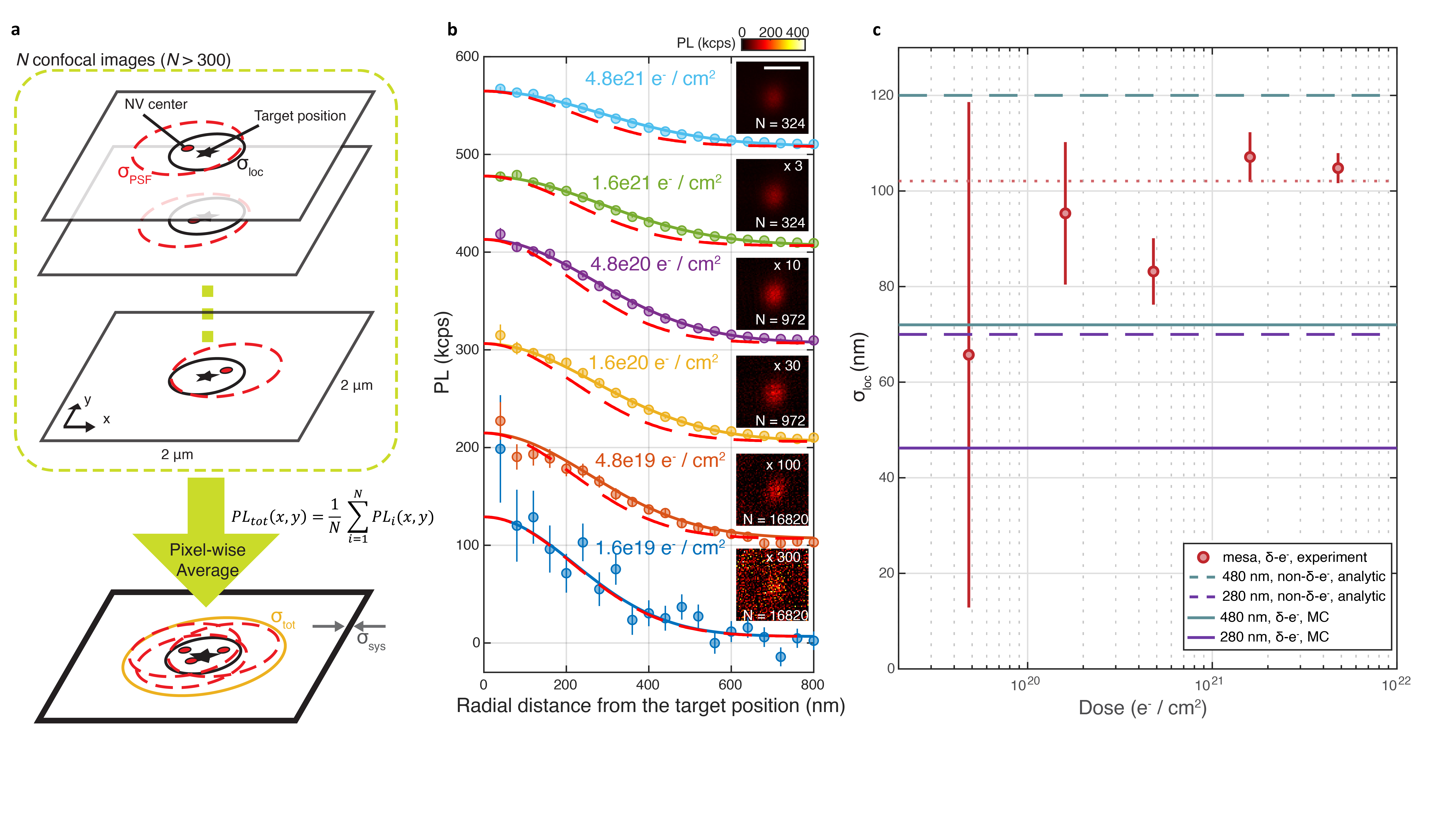}}
    \caption{Quantifying spatial confinement of formed NVs. (a) Schematic of pixel-wise averaging method for estimating \(\sigma_{loc}\). NVs are positioned at the target position (black star) with a lateral precision \(\sigma_{loc}\) (black solid line). Red dashed line indicates the point spread function \(\sigma_{PSF}\) of our confocal microscope. \textit{N} confocal images of unpatterned mesas are pixel-wise averaged to give a \(\sigma_{tot}\) from which \(\sigma_{loc}\) is extracted as discussed in the main text. 
    Residual optical aberrations are indicated by \(\sigma_{sys}\).
    (b) Data points show the radial $PL$ profiles (averaged over angle) of the pixel-wise averaged images
    (bin size: \qty{40}{\nano \meter}). Error bars show standard error of the averaging. 
    Colored solid lines are 2D Gaussian fits, from which \(\sigma_{tot}\) is extracted.
    For comparison, red dashed lines show the radial cuts of the 2D Gaussian functions with a peak width of \( \sqrt{\sigma_{PSF}^2+\sigma_{sys}^2}\).
    Plots are offset for clarity, each with a relative $\Delta PL$ of 100 kcps. 
    Insets show the averaged images with $PL$ scaling inversely proportional to the dose 
    (scale bar: \qty{1}{\micro\meter}). (c) Measured \(\sigma_{loc}\) in the mesas (red circles, error bar: \qty{95}{\percent} confidence interval) is 
    plotted as a function of dose. 
   (the \qty{1.6e19}{\elementarycharge^-\per\centi\meter\squared} dose data point is omitted because of large errorbars \(> \qty{1}{\micro\meter}\)).
   The red dotted line indicates the average \(\sigma_{loc}\) of \qty{102(2)}{\nano\meter} measured in the mesas.
    Solid lines are MC simulations of \(\sigma_{loc}^{pillar}\) in $\delta$-e$^-$-irradiated pillars, 
    which are lower than the analytically calculated \(\sigma_{loc}^{pillar}\) for NVs created without localization methods (dashed lines).   }
    
    \label{fig3}
\end{figure}

We next demonstrate high spatial confinement of NV centers aligned to diamond nanostructures, which is necessary for optimizing spatial overlap between the defect qubit and the structure's photonic modes.
In Figure \ref{fig3}, we quantify the lateral positioning precision of created NVs afforded by $\delta$-electron irradiation. To do so, we first investigate arrays of NVs patterned in \(20\times\)\qty{20}{\micro\meter\squared} mesas and quantify the deviation \(\sigma_{loc}\) of NV positions from their target irradiation spot. The irradiation pattern on the mesa is shown in yellow circles in Figure~\ref{fig1}(c) and example confocal images are shown in Figure \ref{fig2} (bottom right and top left square of each inset). We use these featureless mesas to avoid exciting photonic modes of the nanopillars that modify the NV emission pattern and obfuscate the actual NV position. 
We estimate $\sigma_{loc}$, the standard deviation of lateral NV positioning in the mesas, by pixel-wise averaging 162 tiles in the confocal image, with each \(2\times\)\qty{2}{\micro\meter \squared} tile centered on a single irradiation spot; the tiles were obtained by cutting a \(\sim\)40\(\times \qty{40}{\micro\meter\squared}\) confocal image containing two mesas (the area shown in Figure~\ref{fig1}(c)) into a regular grid (see SI).   
Prior to this image cutting we apply global affine transformations to the original confocal image to account for optical aberrations in our imaging system. 
We then repeat this procedure over several different 40\(\times \qty{40}{\micro\meter\squared}\) confocal image areas
to arrive at the pixel-wise averaged confocal images shown in the insets of Figure~\ref{fig3}(b). 
Each of these images has a finite lateral spread $\sigma_{tot}$ with respect to the target positions due to three contributions: the lateral NV positioning precision of our patterning technique \(\sigma_{loc}\), the point spread function (PSF) of our imaging system \(\sigma_{PSF}\), and residual global aberrations not removed after the first set of global affine transformations \(\sigma_{sys}\):

\begin{equation}
    \sigma_{tot}^2 = \sigma_{loc}^2 + \sigma_{PSF}^2 + \sigma_{sys}^2.
    \label{eqsig}
\end{equation}

To extract \(\sigma_{loc}\) we first measure \(\sigma_{tot}\) by fitting the averaged confocal images with a 2D Gaussian curve, 
\begin{equation}
    PL(x,y) = PL_{max} e^{-(x^2+y^2)/(2\sigma_{tot}^2)},
\end{equation}
where \(PL_{max}\) is the maximum $PL$ of the averaged confocal image. The radial profiles of the averaged images (circles) and the fits (solid lines) are shown in Figure \ref{fig3} (b). 
Then, we measure \(\sigma_{PSF} = \qty{235}{\nano\meter}\) by imaging six single NVs in the mesa region, as described in \cite{hughes_two-dimensional_2023}. We characterize the residual global aberration of our transformed images to be \(\sigma_{sys} =\) \qty{41}{\nano\meter} (see SI).
Finally, we extract \(\sigma_{loc}\) for the different irradiation dosages and plot the results in Figure~\ref{fig3}(c) (red circles). The data shows minimal dependence of \(\sigma_{loc}\) on dose, with an average \(\sigma_{loc}\) = \qty{102(2)}{\nano\meter} (red dotted line). 

We attribute the majority of \(\sigma_{loc}\) to vacancy diffusion during annealing, and hence our measured \(\sigma_{loc}\) provides an estimate of \(D_V = \qty{21}{\nano\meter\squared\per\second}\) via comparison with MC simulations (see SI). This number is consistent with the estimated \(D_V = \qty{17(4)}{\nano\meter\squared \per \second}\) from NV number measurements (Figure~\ref{fig2}) and is within the range of values reported in the literature. 
We note that reported values of \(D_V\) show strong sensitivity to experimental conditions including annealing temperature, annealing time, and vacancy creation method  ~\cite{mclellan_patterned_2016,alsid_photoluminescence_2019,acosta_diamonds_2009,onoda_diffusion_2017, ohno_three-dimensional_2014}. For instance, Ref.~\cite{ohno_three-dimensional_2014} measured \(D_V = \qty{6.5}{\nano\meter\squared\per\second}\) at \qty{850}{\celsius}-\qty{30}{\min} annealing and Ref.~\cite{acosta_diamonds_2009} found \(D_V \leq \qty{40}{\nano\meter\squared\per\second}\) at \qty{1050}{\celsius}-\qty{2}{\hour} annealing.
Further, Ref. ~\cite{onoda_diffusion_2017} found \(D_V\) can be enhanced due to transient dynamics during the first few minutes of annealing (\(D_V \sim \qty{300}{\nano\meter\squared\per\second} \) for \qty{2}{\minute} annealing at \qty{1000}{\celsius}). 

We next estimate the lateral NV positioning precision in the nanopillars \(\sigma_{loc}^{pillar}\). We use MC simulations because a direct measurement is challenging due to the effect of the pillars' photonic modes on the confocal images. 
As shown in Figure \ref{fig3} (c), \(\sigma_{loc}^{pillar}\) (solid lines) is smaller than \(\sigma_{loc}\), 
and notably smaller than \(\sigma_{loc}^{pillar}\) for NVs uniformly distributed in a nitrogen layer bounded by the pillar walls (where \(\sigma_{loc}^{pillar} = 1/4 \times\) pillar diameter), as would result from the conventional method of forming pillars after NV formation. In contrast, our technique achieves improved lateral confinement even in pillars with comparable size to \(\sigma_{loc}\), a fact we attribute to vacancy absorption at the pillar sidewalls during annealing. Specifically, we find \(\sigma_{loc}^{pillar}\) =  \qty{46(1)}{\nano\meter} and \qty{72(1)}{\nano \meter} in our $\delta$-electron irradiated \qty{280}{\nano \meter} and \qty{480}{\nano \meter} pillars.
This improved lateral confinement is expected to enhance the coupling to the nanopillar photonic mode, as shown in the next section. 

\subsection*{Spin and optical properties of single NVs in nanopillars}
\begin{figure}
    
    \centering
    \centerline{\includegraphics[width=105mm]{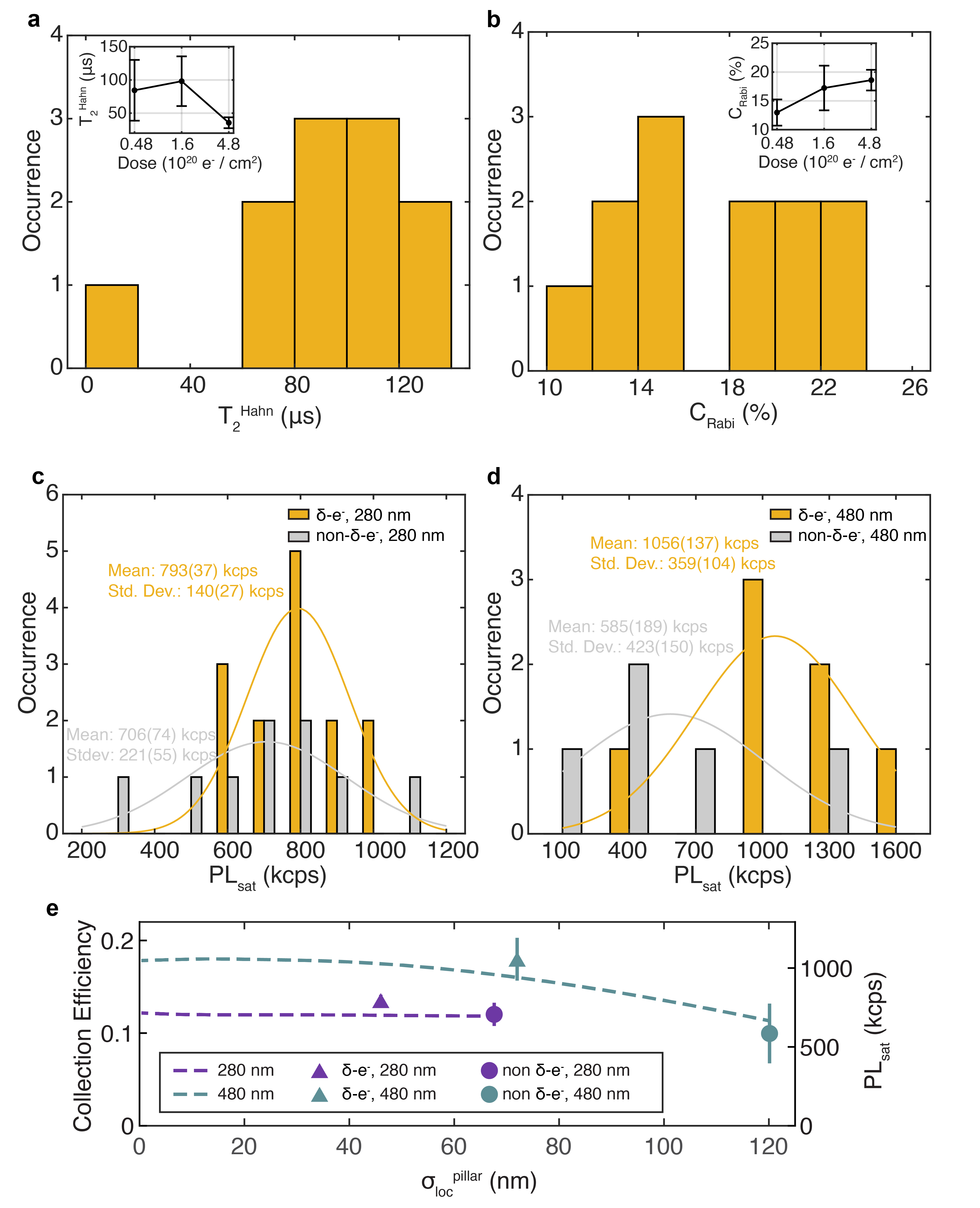}}
    \caption{Spin coherence time \(T_2^{Hahn}\) and $PL$ properties of single NVs formed in nanopillars. (a)  Histogram of \(T_2^{Hahn}\) for NVs $\delta$-electron irradiated with \qty{1.6e20}{\elementarycharge^-\per\centi\meter\squared}. (Inset) Average \(T_2^{Hahn}\) as a function of irradiation dose. (b)  Histogram of Rabi contrast \(C_{Rabi}\) at \qty{1.6e20}{\elementarycharge^-\per\centi\meter\squared}. (Inset) Average \(C_{Rabi}\) as a function of $\delta$-electron irradiation dose. (c-d) Histogram of \(PL_{sat}\) in (c) \qty{280}{\nano\meter} and (d) \qty{480}{\nano \meter} pillars. Each plot shows both non-irradiated (gray) and \qty{1.6e20}{\elementarycharge^-\per\centi\meter\squared}-$\delta$-electron irradiated (yellow) pillars. The solid curves are the Gaussian fit for the histograms. 
    (e) Dashed lines indicate mean photon collection efficiency calculated from FDTD simulations for a given lateral distribution \(\sigma_{loc}^{pillar}\).  
    Data points are the experimentally measured sample mean of the \(PL_{sat}\) distribution for $\delta$-electron irradiated pillars (triangles) and non-irradiated pillars (circles) for \qty{480}{\nano\meter} (teal) and \qty{280}{\nano\meter} (purple) diameter pillars. The error bars denote the standard error of the estimation of the population mean. The limits of the secondary y-axis are chosen so that the simulated collection efficiency and measured $PL_{sat}$ for the nonirradiated pillars line up. 
    } 
    \label{fig4}
\end{figure}

We next present the spin coherence and photoluminescence properties of single $^{15}$NV centers created inside nanopillars.
Figure~\ref{fig4}(a) shows a histogram of the Hahn echo coherence time, \(T_2^{Hahn}\), of 12 NV centers formed by \qty{1.6e20}{\elementarycharge^-\per\centi\meter\squared}-irradiation.
We observe reliably long ${T}^{Hahn}_2$ with a mean of \qty{98 (37)}{\micro\second}, which we attribute to the gentle nature of our NV formation process, producing little collateral damage that can adversely affect coherence. 
The coherence time is consistent with the limit imposed by the surrounding substitutional nitrogen (P1 center) bath(see SI).
In contrast, \qty{15}{keV} ion implantation results in NVs with the majority \((>\qty{90}{\percent})\) having ${T}^{{Hahn}}_2$ less than \qty{50}{\micro\second}\cite{wang_self-aligned_2022}.
As shown in the inset, we also measure long ${T}^{{Hahn}}_2$ for single NV pillars with \qty{4.8e19}{\elementarycharge^-\per\centi\meter\squared} irradiation, while we see a reduced ${T}^{{Hahn}}_2$ of \qty{36(8)}{\micro\second} for \qty{4.8e20}{\elementarycharge^-\per\centi\meter\squared} (Figure~\ref{fig4}(a), inset). We attribute this reduction to 
increased vacancy-related damage, consistent with previous reports \cite{alsid_photoluminescence_2019,hughes_two-dimensional_2023}.

We also demonstrate favorable optical properties of our $\delta$-electron irradiated single NV centers, namely good spin-dependent optical readout contrast and high photon collection rates. Fig. \ref{fig4}(b) shows an average spin-dependent Rabi $PL$ contrast \(C_{Rabi}= \frac{PL_0-PL_{\pm{1}}}{PL_0}\) of \qty{18(4)}{\percent} at \qty{1.6e20}{\elementarycharge^- \per\centi\meter\squared}, where \(PL_0\) and \(PL_{\pm 1} \) are the $PL$ for NV electronic spin states 0 and \(\pm 1\), respectively . The inset shows the dependence of \(C_{Rabi}\) on irradiation dose with no evidence of reduced contrast at the highest dosages compatible with single NV formation.

In Figure \ref{fig4}(c) and (d), we show histograms (yellow) of the saturation count rate \(PL_{sat}\) for \qty{280}{\nano\meter} and \qty{480}{\nano\meter} pillars, respectively, where \(PL_{sat}\) is measured as a function of \qty{532}{\nano\meter} excitation power \(P_{exc}\) (see SI). 
The means of the two \(PL_{sat}\) histograms are 0.793(37) Mcps and 1.056(137) Mcps, respectively.
Also plotted in grey are the \(PL_{sat}\) histograms of non-irradiated pillars with as-grown single \(^{15}\)NVs, which have a uniform spatial distribution inside the pillars. The non-irradiated pillars show a lower mean \(PL_{sat}\) compared to the $\delta$-electron irradiated pillars by a factor of 1.8 and 1.1 for \qty{480}{\nano\meter} and \qty{280}{\nano\meter} pillars, respectively.
The increase in mean $PL_{sat}$ in the $\delta$-electron irradiated 480 nm pillars is statistically significant. 

In Figure~\ref{fig4}(e), we conduct finite-difference time-domain (FDTD) simulations to study the effect of lateral localization on $PL_{sat}$ in nanopillars (see Methods). The simulations (dashed lines) and the data are in good agreement, indicating that increased localization precision is the main contributor to increased \(PL_{sat}\) for the \qty{480}{\nano\meter} pillars, while \(PL_{sat}\) does not depend strongly on \(\sigma_{loc}^{pillar}\) in the 280 nm pillars. For both pillars, our \(\sigma_{loc}^{pillar}\) nearly maximizes photon collection efficiency.
For future applications, color center localization should be performed in conjunction with FDTD simulations to optimize the emitter overlap with the photonic mode. 
Overall, high collection efficiency combined with long coherence and large \(C_{Rabi}\) shown here is crucial for realizing advanced functionalities in devices for NV-based sensing, as discussed in the next section.

\subsection*{High-yield, scalable magnetic field sensors}
Lastly, we present an outlook for the improvements our method offers to scalable, high-yield fabrication of highly sensitive magnetic field sensors.  

In a typical optical spin-state readout scheme, the alternating current (AC) magnetic field sensitivity \(\eta\) is given as
\begin{equation}
\eta = \frac{ \hbar}{  g_e \mu_B} \frac{1}{e^{(-2\tau/T_2^{Hahn})} \sqrt{2\tau}} \sqrt{1+\frac{4}{C_{Rabi}^2 n_{\mathrm{avg}}}},
\label{eqmagsens}
\end{equation}
where  \(\hbar\) is the reduced Planck constant, \(g_e \approx 2\) is the NV's electronic g factor, \(\mu_B\) is the Bohr magneton, \(2\tau\) is the total free evolution time, and \(n_{\mathrm{avg}}\) is the average photon number per measurement. This expression highlights the importance of long $T_2^{Hahn}$, large $C_{Rabi}$, and high $PL$ for sensing small magnetic fields.

\begin{figure}
    
    \centering
    \centerline{\includegraphics[width= 80mm]{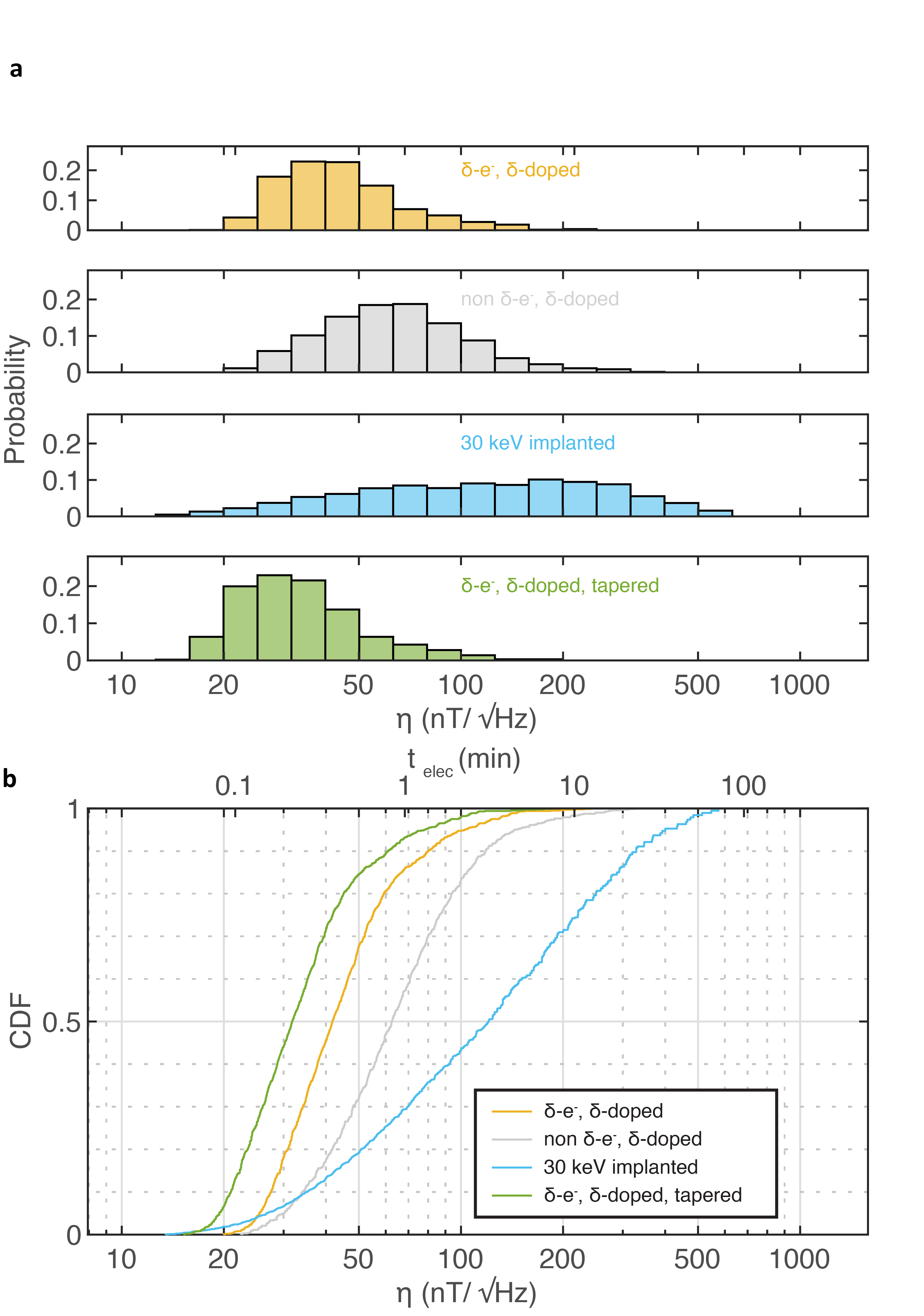}}
    \caption{High-yield fabrication of highly sensitive magnetic field sensors. (a) Histogram and (b) cumulative density function (CDF)  of the estimated AC magnetic field sensitivity \(\eta\) of single NVs in \qty{480}{\nano\meter} pillars. The \(\eta\) distribution is estimated for $\delta$-doped, $\delta$-electron irradiated (yellow) and $\delta$-doped, non-irradiated (gray) pillars using measured \(PL_{sat}\), \(T_{2}^{Hahn}\) and \(C_{Rabi}\) distributions. The \(\eta\) distribution of conventionally implanted pillars (cyan) is generated using \(PL_{sat}\) and \(C_{Rabi}\) measurements on our non-irradiated pillars with reported \(T_2^{Hahn}\) distribution for \qty{30}{keV} implantation \cite{jakobi_efficient_2016}.  The distribution for $\delta$-doped, $\delta$-electron irradiated pillar with better sidewall taper angle of \(\qty{70}{\degree}\) (green) is also estimated using  \(T_{2}^{Hahn}\) and \(C_{Rabi}\) measurements on our $\delta$-electron irradiated pillars with the estimated $PL$ improvement from FDTD simulations. A secondary x-axis shows the minimum averaging time for a \qty{53}{\nano\meter}-deep NV to detect a single electron spin located on the diamond surface.}
    \label{fig5}
\end{figure}

Figure \ref{fig5}(a) shows simulated histograms of \(\eta\) for NVs in pillars formed via our method (yellow) compared to two other methods: conventional 30 keV nitrogen-implanted layers (cyan) and $\delta$-doped layers without lateral localization (gray). Also plotted are $\delta$-electron irradiated pillars with future improvements to the pillar geometry (green). 
The distributions are generated using Equation~\ref{eqmagsens} with \(n_{\mathrm{avg}} = 0.5\cdot PL_{sat} \cdot\qty{400}{\nano\second}\) and \(2\tau = T_2^{Hahn} \) with the \(PL_{sat}\), \(C_{Rabi}\) and \(T_{2}^{Hahn}\) distributions experimentally measured in this work.
From this histogram, we calculate the cumulative density function \( P(X < \eta) \) in  Figure \ref{fig5} (b). The median \(\eta\) of NVs formed using our method is \qty{42}{\nano\tesla / \sqrt \hertz} with \qty{86}{\percent} of the NV centers exhibiting \(\eta\) \(\lesssim \qty{68}{\nano\tesla / \sqrt \hertz}\). For reference, a 53-nm deep NV with \(\eta\) \(= \qty{68}{\nano\tesla / \sqrt \hertz}\) can detect a single electron spin at the diamond surface in a typical averaging time of 1 minute. For the non-localized $\delta$-doped method, the median \(\eta\) is \qty{63}{\nano\tesla / \sqrt \hertz} with \qty{57}{\percent} of the NVs exhibiting \(\eta\) \(\lesssim \qty{68}{\nano\tesla / \sqrt \hertz}\), where we use the measured \(PL_{sat}\), \(C_{Rabi}\), and \(T_{2}^{Hahn}\) distributions of non-irradiated pillars. With the conventional, implantation-based method, the median \(\eta\) is \qty{121}{\nano\tesla / \sqrt \hertz} with \qty{29}{\percent} of the NVs exhibiting \(\eta\) \(\lesssim \qty{68}{\nano\tesla / \sqrt \hertz}\), where we use the measured \(PL_{sat}\) and \(C_{Rabi}\) distributions from non-irradiated pillars with the reported distribution of \(T_{2}^{Hahn}\) from \qty{30}{keV} implantation\cite{jakobi_efficient_2016}, chosen because it produces a similar NV depth of 40-\qty{50}{\nano\meter}. Hence our method produces a significantly higher yield of high-sensitivity NV magnetometers, where we demonstrate an estimated 3-fold higher yield of single-electron-spin detectable magnetometers compared to conventional implantation-based methods. 
We note that further improvements can be realized by utilizing pillars with a \( \qty{70}{\degree}\) sidewall taper angle\cite{wang_self-aligned_2022}, as shown in green in Figure \ref{fig5}. 
Higher-order dynamical decoupling can also extend coherence time, where an order of magnitude increase for shallow, $\delta$-doped NVs has been demonstrated \cite{myers_probing_2014}, leading to a further 3\(\times\) improvement in \(\eta\) to \(< \qty{10}{\nano\tesla / \sqrt \hertz}\).

\subsection*{Conclusions}
To conclude, we demonstrate three-dimensional localized formation of highly coherent NV centers aligned to prefabricated nanophotonic structures. Using our method, we find NV spin and photoluminescence properties superior to those for NV centers formed via conventional implantation methods as well as nonlocalized $\delta$-doped methods. These improved properties culminate in a significantly higher yield of high-sensitivity magnetometers, an important application of NV centers. Through our work, we also gain an understanding of vacancy diffusion in nanostructured diamond devices.

While we demonstrate our technique here on NV centers in diamond nanopillars, we emphasize that our method can be readily applied to other device geometries, such as 2D and 1D photonic crystal cavities \cite{schroder_scalable_2017,hausmann_coupling_2013} and nano-optomechanical devices \cite{cady_diamond_2019}
as well as to other material systems,
including divacancies\cite{christle_isolated_2015} and silicon vacancies\cite{widmann_coherent_2015} in silicon carbide and T centers in silicon\cite{macquarrie_generating_2021}. In these optically addressable qubit systems, 
the advantages outlined here can be transferred using a similar targeted irradiation technique guided by our model of vacancy diffusion and capture for different geometries.

Looking forward, there is still room for further improvement in the collective control over the number, position, and coherence of color centers. The ultimate goal of forming a single defect with unit probability at a spot can be achieved by, for instance, $\delta$-electron irradiation with in-situ annealing and photoluminescence characterization. 
More accurate positioning can be achieved by reducing the annealing time and optimizing irradiation parameters (\textit{e.g.}, smaller spot size). 
Overall, our results strengthen the role of optically addressable solid-state spin defects in next-generation metrology and information science.

\section*{Methods}
\subsection*{PECVD diamond growth}
Diamond homoepitaxial growth and $\delta$ doping were performed via plasma-enhanced chemical vapor deposition (PECVD) using a SEKI SDS6300 reactor on a (100) oriented electronic grade diamond substrate (Element Six Ltd.). Prior to growth, the substrate was fine-polished by Syntek Ltd. to a surface roughness of \(\sim\)200-300 pm, followed by a 4-\qty{5}{\micro \meter} etch to relieve polishing-induced strain. 
The growth conditions consisted of a 750 W plasma containing 0.1$\%$ $^{12}$CH$_{4}$ in 400 sccm H$_2$ flow held at 25 torr and $\sim$730 $^{\circ}$C according to a pyrometer.
A $\sim$154 nm-thick isotopically purified (99.998$\%$ $^{12}$C) epilayer was grown. During the nitrogen $\delta$-doping period of growth, 
$^{15}$N$_2$ gas
\qty{1.0}{\percent} of the total gas content) is introduced into the chamber for five minutes. After growth, the sample was characterized with secondary ion mass spectrometry (SIMS) to estimate the isotopic purity, epilayer thickness, and properties of the $\delta$-doped layer (3.6 nm thick, 98 ppm*nm, see SI). 

\subsection*{Monte-Carlo simulations}
We simulate NV center formation using Monte Carlo (MC) simulation. Our simulation models the dominant effects during the annealing step, namely the diffusion of monovacancies within the prefabricated device and their capture by the existing nitrogen atoms to form NV centers. We also consider vacancy recombination at the diamond surfaces \cite{pezzagna_creation_2010, racke_vacancy_2021} (\textit{e.g.}, the top surface and the pillar's sidewalls for nanopillars).

Simulating the atomic-scale diffusion process on a diamond lattice is computationally intensive, necessitating approximately 50 million discrete "jumps" for a vacancy to traverse \qty{1}{\micro\meter}, with each jump spanning a minuscule \qty{0.154}{\nano\meter} C-C bond spacing. Therefore, we adopt a coarse-grained approach with a cubic lattice of spacing \qty{1}{\nano\meter}, still significantly smaller than the device dimensions. Within each simulation run, we randomly select positions for $N_N$ nitrogen atoms within the $\delta$-doped layer region and \(N_V\) vacancies within the vacancy-rich area from \qty{200}{\kilo\electronvolt} electron irradiation. We estimate \(N_V\) from CASINO simulations with a scaling factor \(\alpha\), which is set as a free parameter (see SI). 
For computational efficiency, we only consider vacancies with depth ($<$\qty{1}{\micro\meter}) since deeper vacancies do not contribute to NV center formation.

Given the initial conditions of the simulations, we segment the annealing process into shorter time steps. In each step, all vacancies randomly move some number of jumps, after which we check if they encountered a capture event. A capture by a nitrogen atom can occur when a monovacancy is in the same coarse-grained cell as a nitrogen atom with a probability of \(\frac{16V_{cc}/V_{uc}}{(8V_{cc}/V_{uc})^2/2}\), where \(V_{cc}\) and \(V_{uc}\) are the volumes of the coarse-grained cell and unit cell, respectively. When an NV center forms, both the vacancy and the nitrogen atom are removed from the simulation during the subsequent time steps. Conversely, if a monovacancy gets captured by the boundaries, only the vacancy is eliminated. 


\subsection*{Finite-difference time-domain simulations}
Lumerical FDTD software is used to simulate the collection efficiency of photons emitted by NVs inside nanopillars. We model an NV as a point source consisting of two orthogonal dipoles perpendicular to the NV axis. The emission frequency range of the dipole is chosen to match the frequency range of the phonon sideband of the NV emission spectrum at room temperature: \(\qty{650}{}-\qty{800}{\nano\meter}\). The NVs are positioned \qty{53}{\nano\meter} below the top surface. We simulate our two nanopillar geometries with diameters \qty{280}{\nano\meter} and \qty{480}{\nano\meter} with the side wall angle of \(\sim \qty{83}{\degree}\) and height of \qty{1.4}{\micro\meter}, which are attached to a diamond slab of finite thickness. For computational efficiency, we set the thickness of the slab to be \qty{1}{\micro\meter}. To avoid any interference due to this relatively thin slab, we absorb all incoming fields at the bottom surface of the slab. We do this by setting the simulation area such that the bottom diamond interface matches the perfectly absorbing simulation boundary. The collection efficiency is then calculated from the power transmitted to a monitor plane just below the pillar inside the slab. We calculate the far-field emission through a collection cone with \(\text{NA} = 0.7\).

To calculate the mean collection efficiency for the distribution of NVs with a given \(\sigma_{loc}^{pillar}\), we first sweep the position of the NV laterally in two orthogonal directions ($dx$ and $dy$) and calculate the collection efficiency. 
Then, we extrapolate the collection efficiency for a given radial displacement $\vec{dr} = (dr \cos{\theta},dr \sin{\theta})$ by assuming the superposition of two orthogonal NVs. In particular, the collection efficiency is calculated as a weighted average of those calculated at two orthogonal displacements $dx=dr$ and $dy=dr$, where the weights are given as \(\cos^2{\theta}\) and \(\sin^2{\theta}\), respectively. 
We sweep the lateral confinement of NVs from perfectly localized (\(\sigma_{loc}^{pillar} = \qty{0}{\nano\meter}\)) to maximally delocalized (\(\sigma_{loc}^{pillar} = 1/4 \times\) pillar diameter, corresponding to a uniform lateral distribution across the pillar). For simplicity, we set the lateral probability distribution to follow a 2D Gaussian function with a spread of \(\sigma_0\) which we truncate at the pillar boundary beyond which the probability is zero. For a given \(\sigma_0\), we use the probability distribution and simulated collection efficiency to calculate both \(\sigma_{loc}^{pillar}\) and the mean of the collection efficiency.

\clearpage

\clearpage

\clearpage

\clearpage

\clearpage

\section{Acknowledgments}
The authors thank Hitoshi Kato and JEOL Ltd. for \qty{200}{\kilo\electronvolt} irradiation. 
We gratefully acknowledge the support of the Gordon and Betty Moore Foundation’s EPiQS Initiative via Grant GBMF10279. We also acknowledge support from Cooperative Research on Quantum Technology (2022M3K4A1094777) through the National Research Foundation of Korea(NRF) funded by the Korean government (Ministry of Science and ICT(MSIT)), and from the DOE Q-NEXT Center (Grant No. DOE 1F-
60579).
The authors acknowledge the use of shared facilities of the UCSB Quantum Foundry through Q-AMASE-i program (NSF DMR1906325), the UCSB MRSEC (NSF DMR 1720256), the UCSB Nanofabrication Facility (an open access laboratory), and the Quantum Structures Facility within the UCSB California NanoSystems Institute. S. A. M and  acknowledges support from UCSB Quantum Foundry. L. B. H. acknowledges support from the NSF Graduate Research Fellowship Program (DGE 2139319) and the UCSB Quantum Foundry.

\bibliography{references.bib}

\end{document}


\title{Scalable, nanoscale positioning of highly coherent color centers in prefabricated diamond nanostructures }

\maketitle

\section{NV number estimation via maximum likelihood estimation}\label{suppl_MLE}
To estimate the average number of nitrogen vacancy (NV) centers per pillar, we measure continuous wave-electron spin resonance (CW-ESR) spectra of 121 pillars each for a given pillar size, irradiation dose, and annealing condition. 
In a given set of 121 pillars, we count the number of NV orientations \(l\) for each pillar to measure the probability distribution \(P^{exp}_{\lambda}(l)\), where \(\lambda\) is an average number of NV per pillar to be estimated. 

\begin{figure}
    \centering
    \centerline{\includegraphics[width= 180mm]{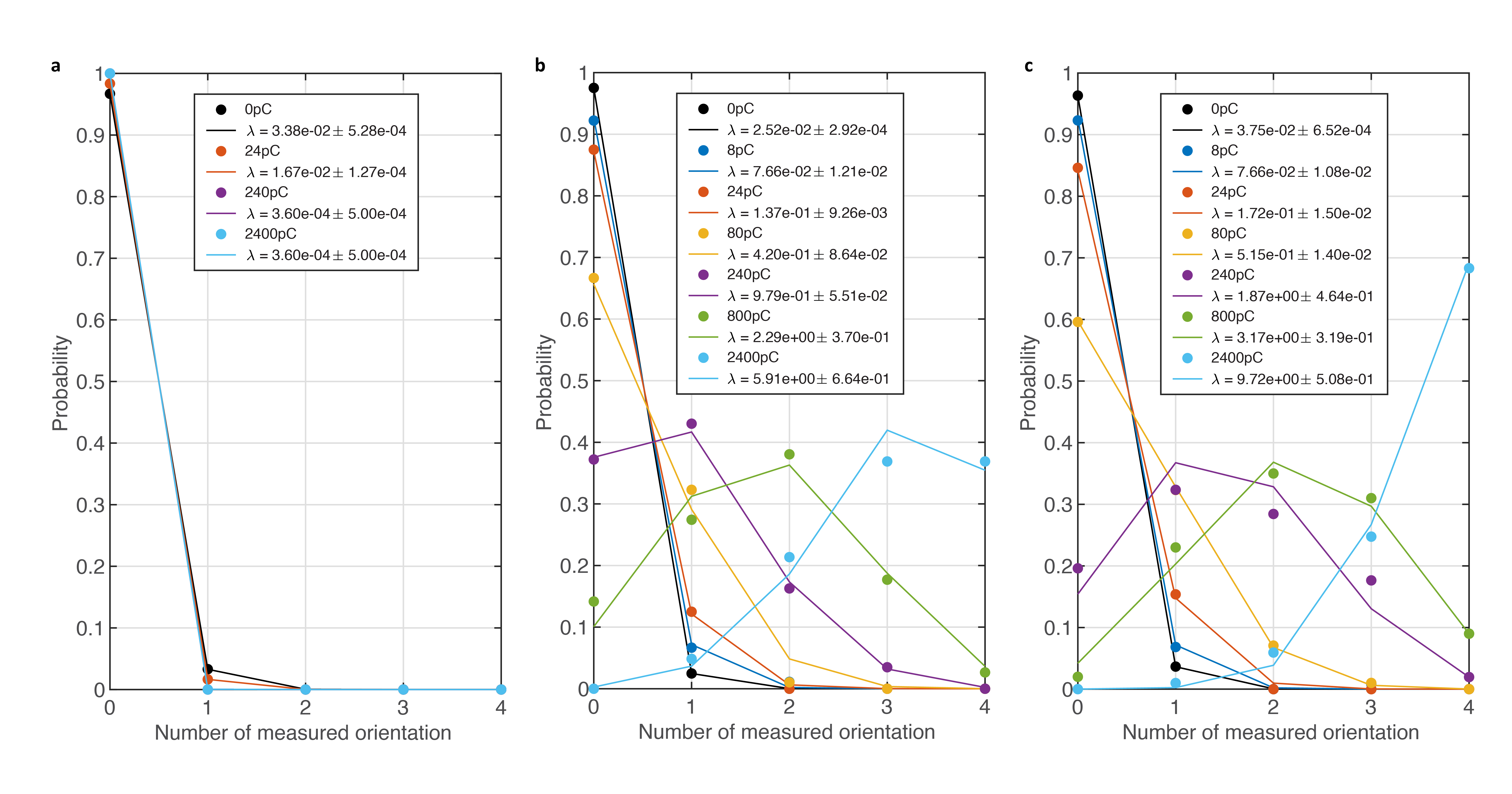}}
    \caption{Probability distribution of measured NV orientations in a nanopillar. We measure probability distribution (filled circles) of a) non-annealed \qty{480}{\nano\meter} pillars, b) annealed \qty{280}{\nano\meter} pillars, and c) annealed \qty{480}{\nano\meter} pillars for different irradiation doses. Each plot is a result of measuring 121 pillars. We perform MLE to find the best fit to the model probability distribution (solid lines) parameterized by $\lambda$, the mean number of NVs per pillar. Indicated errors on $\lambda$ denote \qty{95}{\percent} confidence intervals of the fit. }
    \label{figSI_MLE}
\end{figure}

In Figure~\ref{figSI_MLE}, we perform maximum likelihood estimation (MLE) to fit \(P^{exp}_{\lambda}(l)\) to a model distribution \(P_{\lambda}(l)\) and extract  \(\lambda\) \cite{mclellan_patterned_2016}. We model the probability distribution as
\begin{equation}
    P_{\lambda}(l) = \sum_{n = 0}^{\infty} \frac{\lambda^n e^{-\lambda}}{n!} (1/4)^n S(n,l) {}_4P_l ,
\end{equation}
where \(S(n,l)\) is the Stirling number of the second kind and \({}_4P_l\) is the number of permutations of \(l\) orientations from a total of 4 orientations. This distribution assumes that the number of NVs per pillar follows a Poisson distribution. 
Also, the units for irradiation dose are in total number of electrons, where \qty{1}{\pico\coulomb} = \qty{2e18}{\elementarycharge\per\centi\meter\squared}.

Note that the MLE method becomes less precise for very low (\(\lambda <<1 \)) or large (\(\lambda >> 16\)) average NV number per pillar.
We characterize the additional uncertainty due to the systematics of our method by numerically simulating 100 sets of 121 random variables following a Poisson distribution with a given expectation \(\lambda_{true}\). Then, we perform MLE for each set \(i\) to get \(\lambda_i\), from which we characterize the systematic uncertainty by calculating the standard deviation of the error \(\sqrt{ \frac{1}{100} \sum_{i = 1}^{100} {(\lambda_i-\lambda_{true})^2}}\). The additional error is added in Figure~2 in the main text.

\section{Secondary Ion Mass Spectrometry}
\begin{figure}
    \centering
    \includegraphics[width= 100mm]{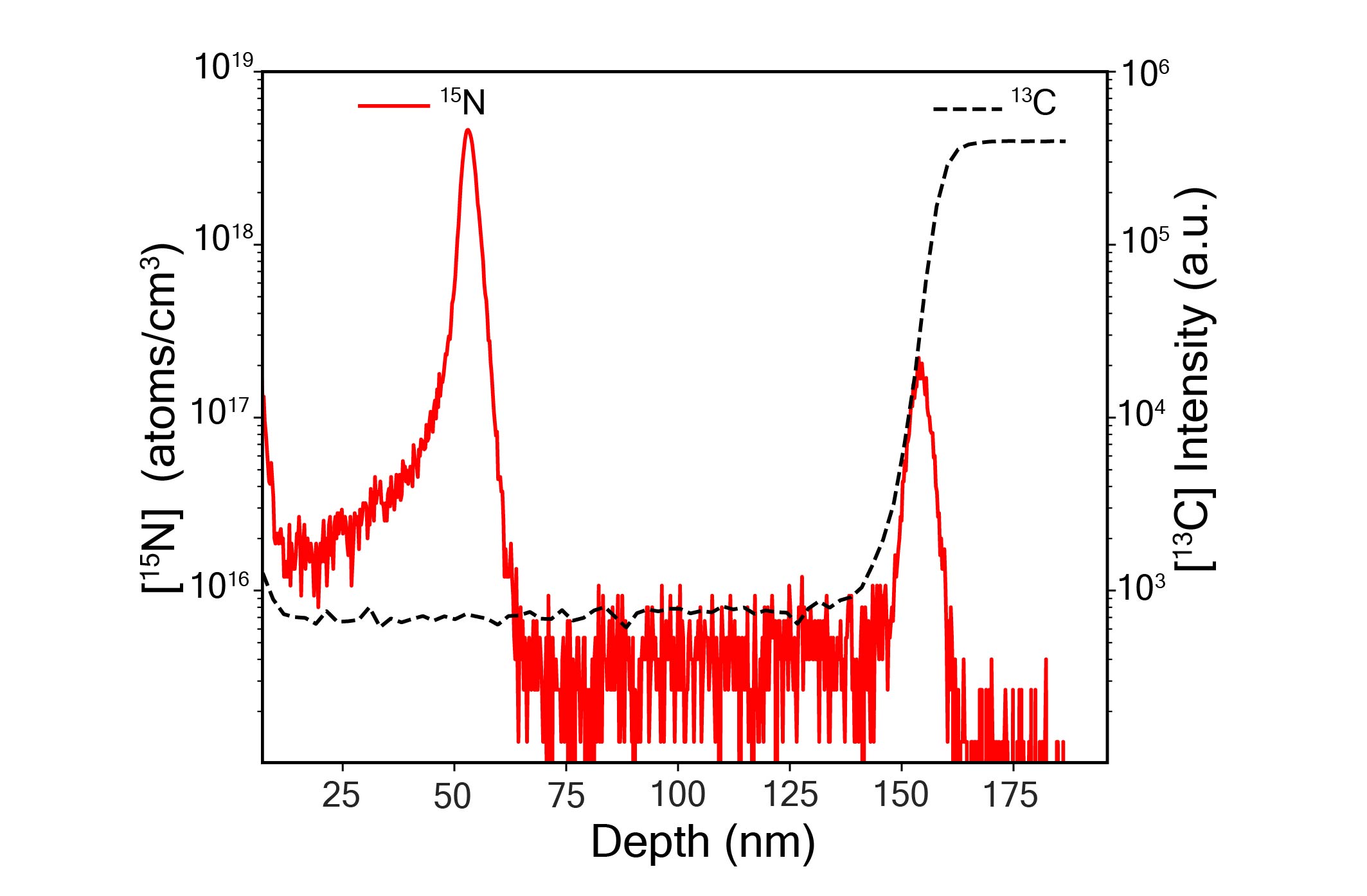}
    \caption{SIMS depth profile showing the $^{15}$N (collected as a $^{12}$C$^{15}$N$^-$ ion) and $^{13}$C composition as a function of depth. 
    }
    \label{figS_SIMS}
\end{figure}

SIMS is performed with a CAMECA IMS 7f dynamic instrument using a primary Cs$^{+}$ beam energy of 7 kV and current of $\sim$30 nA at an incident angle of $\SI{21.7}{\degree}$. 
The sample is biased to -3000 V and $^{12}$C$^{15}$N$^{-}$ negative secondary ions are detected using a high mass resolving power, M/$\Delta$M = 6006. 
Only ions from the central $\SI{33}{\micro\meter}$ are collected from the $\SI{100}{\micro\meter}$ sputtering crater to avoid edge effects.
Fig.~\ref{figS_SIMS} shows a SIMS depth profile of the sample studied in this work, highlighting the composition of $^{15}$N (collected as a $^{12}$C$^{15}$N$^-$ ion, displayed in red) and $^{13}$C (black, dashed).
The drop in $^{13}$C concentration indicates the start of the $^{12}$C purified epitaxy, and an unintentional nitrogen peak is seen at this substrate-nitrogen interface.
The intentionally $\delta$-doped layer occurs at 53 nm deep and has a thickness of 3.66(2) nm, as determined from the full width at half-maximum (FWHM) of a Gaussian fit to the peak.
To calculate the areal density in the 53-nm-deep peak, the peak is integrated over three standard deviations, resulting in 1.736(9)$\times10^{12}$ atoms/cm$^{2}$ or 98.6(5) ppm$\cdot$nm. 
The stated errors represent a 95\% confidence interval for a Gaussian peak fit. 

To ensure that our results presented in the main text are dominated by the NVs in the \qty{53}{\nano\meter}-deep peak, and not from the interface peak at a depth of 154 nm, we similarly calculate the $^{15}N$ areal density in the 154-nm-deep peak and find it to be 5.2 ppm$\cdot$nm. 
Hence \qty{94}{\percent} of the $^{15}N$ is in the intentionally-doped 53-nm-deep peak. We also note that when selecting NV centers for measurement, we rule out those which show $^{13}$C bath coupling in a Hahn-echo sequence\cite{childress_coherent_2006}. For the $T_2$ data presented in Figure 4, 2 out of 25 NV centers measured displayed $^{13}$C bath coupling.  

\section{Vacancy formation via $\delta$-electron irradiation}
We use a \qty{200}{keV} electron beam with \qty{20}{\nano\meter} diameter spot size ($\delta$-electron irradiation) to create vacancies at the center of the nanopillars. Alignment marks fabricated on the sample allow navigation of the electron beam to the target positions. The alignment accuracy and minimal beam size of the electron irradiation confines the irradiation-induced damaged region of the diamond to the localized position with respect to the prefabricated nanostructure. When its energy exceeds \qty{145}{keV}, an electron can create a Frenkel defect in the diamond\cite{mclellan_patterned_2016}. A Frenkel defect is created by a carbon atom displaced from its original lattice site to create a vacancy site and an interstitial carbon site.

\begin{figure}
    \centering
    \centerline{\includegraphics[width= 100mm]{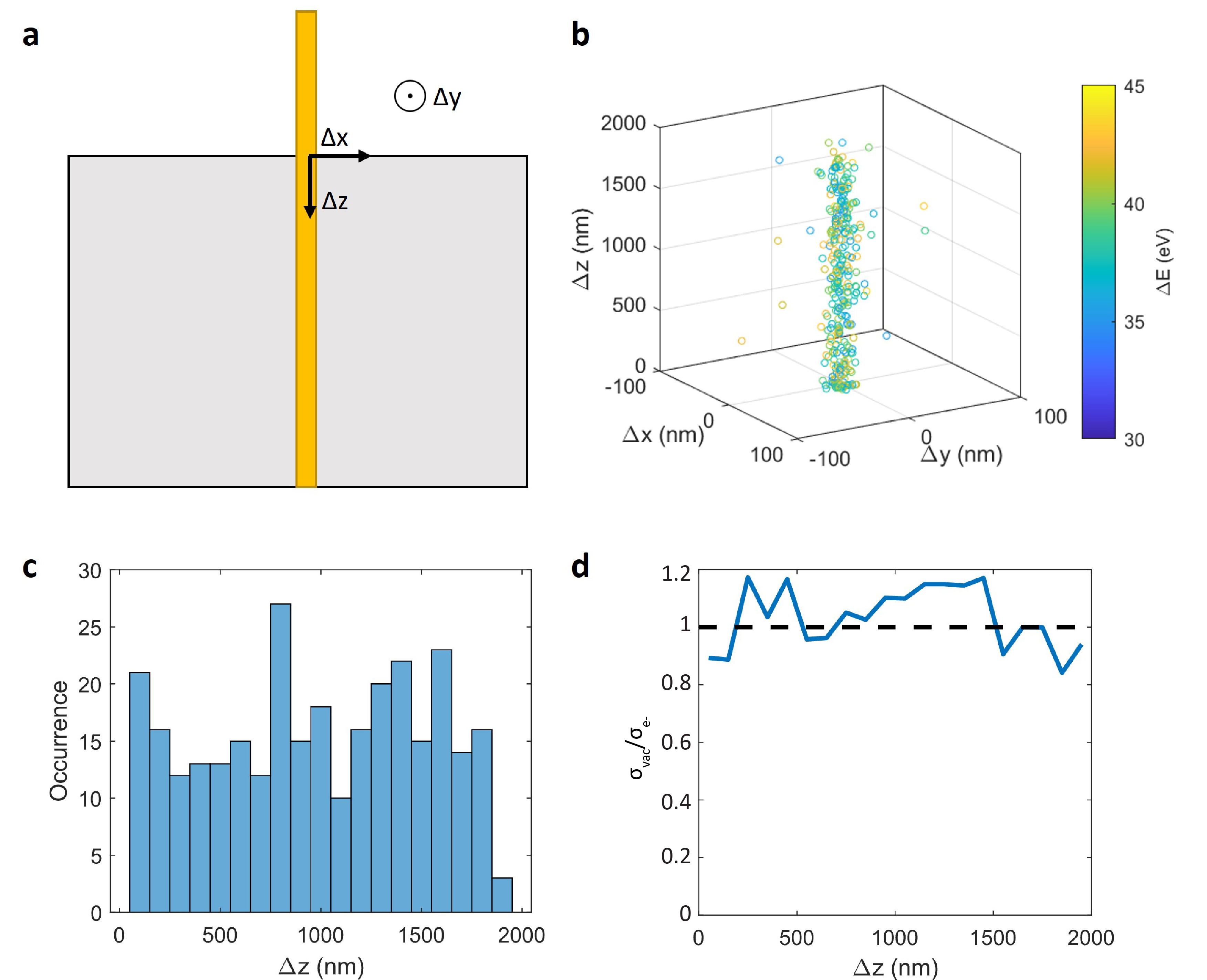}}
    \caption{CASINO simulation of vacancy creation. (a) Schematic of CASINO simulation. The \qty{200}{keV} electron beam with a spot size of \qty{20}{\nano\meter} (yellow) is incident on the top surface of a \qty{2}{\micro\meter} thick diamond sample (gray). 
    (b) Vacancy creation sites where an electron scatters from a carbon atom and loses kinetic energy of \(\Delta E\) greater than the displacement energy \(E_d = \qty{35}{eV}\). The colors of the markers indicate \(\Delta E\). (c) Histogram of the depths of the vacancy sites. (d) The median lateral displacement of the vacancies \(\sigma_{vac}\) as a function of \(\Delta z\). \(\sigma_{vac}\) is normalized to the size of the electron beam \(\sigma_{e^-}\), where the normalized value of 1 is highlighted with the dotted line . }
    \label{figS_CASINO}
\end{figure}

We implement CASINO simulations\cite{drouin_casino_2007} to characterize vacancy formation via $\delta$-electron irradiation, as shown in Figure \ref{figS_CASINO}(a). We generate \qty{2000000}{} independent trajectories of electrons with the same energy and spot size as our electron beam in a \qty{2}{\micro \meter} thick diamond. Then, we locate the scattering sites where the energy loss of the electron \(\Delta E\) due to scattering is greater than the threshold displacement energy \(E_d = \qty{35}{eV}\) of diamond  \cite{bourgoin_threshold_1976}; we consider these sites as vacancy sites. In Figure \ref{figS_CASINO}(b), a total of 340 vacancy sites have been formed, corresponding to \qty{8.5e-5}{vacancy \per \micro\meter} electron. 

The number of simulated vacancy sites has minimal dependence on depth for the entire range of \qty{2}{\micro\meter}, as shown in Figure \ref{figS_CASINO}(c) and consistent with ref~\cite{campbell_radiation_2000}, where vacancy density is depth independent up to a few tens of \qty{}{\micro\meter}. 
Another important feature of our simulation is that the vacancy sites remain laterally confined to the spot size throughout the \qty{2}{\micro\meter} depth extent of the simulated diamond, as shown in Figs \ref{figS_CASINO}(d). This narrow electron trajectory is a result of the vertical momentum of electrons being sufficiently high such that any collisions cannot effectively divert the trajectories of the electrons up to the depth of our simulation.

One can estimate the number and spatial distribution of monovacancies using the simulation with several caveats. First, interstitial carbon atoms form in the same amount as the number of vacancies, where they can recombine during annealing. 
In another effect, a carbon atom displaced by electron irradiation can create additional vacancies along the trajectory. However, we expect this latter effect to be minimal since the maximum possible energy transfer from a \qty{200}{keV} electron to a carbon atom is lower than the minimum kinetic energy for a carbon atom to initiate such a cascade effect \cite{losero_creation_2023}.
Thus, the density of monovacancies is expected to be lower than what we simulate.

\section{Modeling NV formation}
The number of created NVs \(N_{NV}\) estimated from MC simulations for a given $\delta$-electron irradiation depends on the number of monovacancies \(N_V\), the number of nitrogen \(N_N\) within the monovacancy diffusion area, and the probability of capture when a vacancy migrates to a neighboring site of a nitrogen atom \(p_{cap}\). 
However, while \(N_N\) and \(N_{NV}\) and can be experimentally measured, \(N_V\) and \(p_{cap}\) are hard to measure and thus considered free parameters to fit the data.
Furthermore, when \(N_V\) is sufficiently small, \(N_{NV} \propto N_N \cdot p_{cap}\cdot N_V\), which makes it difficult to distinguish their contributions independently.
Hence, we define  \( p_{cap} N_V \equiv \alpha N_V^{max} \), where \(\alpha\) is a constant prefactor and \(N_V^{max}\) is the maximum number of vacancies calculated from the CASINO simulation.

\begin{figure}
    \centering
    \centerline{\includegraphics[width= 180mm]{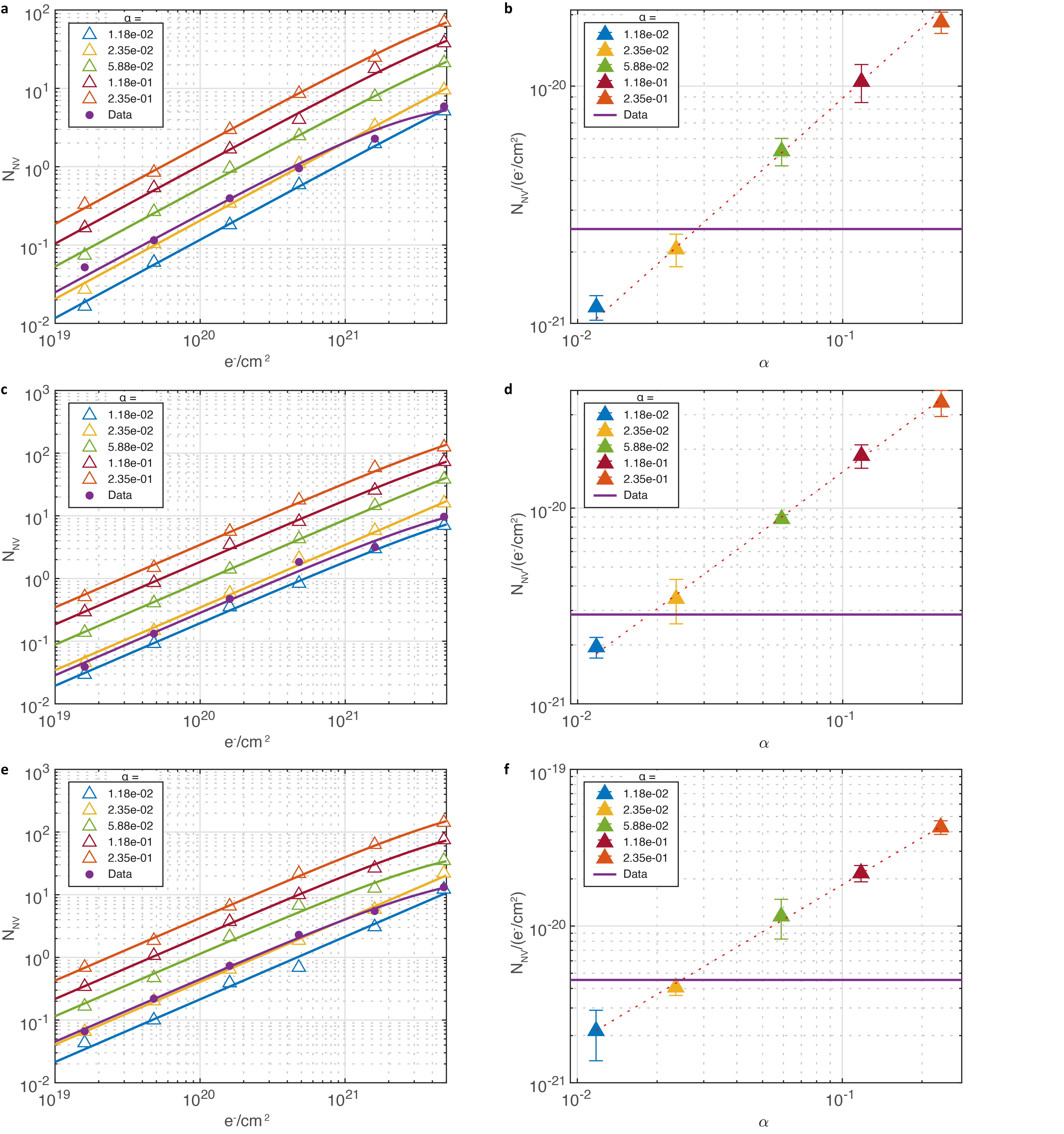}}
    \caption{Estimating \(\alpha\). Simulated $N_{NV}$ as a function of \(\delta\)-electron irradiation dose and \(\alpha\) for (a) \qty{280}{\nano\meter} pillar, (c) \qty{480}{\nano\meter} pillar and (e) mesa (empty triangles). Our measured \(N_{NV}\) for each geometry are shown in purple circles. We fit all data with a function with a linear term and a saturation term\cite{alsid_photoluminescence_2019}. The extracted linear terms for the simulation are shown for (b) \qty{280}{\nano\meter} pillar, (d) \qty{480}{\nano\meter} pillar and (f) mesa (filled triangles, errorbars: \qty{95}{\percent} confidence interval). The linear terms are proportional to \(\alpha\), where the mean prefactors for each geometry are plotted in red dotted lines. We use this value to extract \(\alpha\) of 0.024(2). }
    \label{figSI_alpha}
    \end{figure}

In Figure~\ref{figSI_alpha}, we characterize the prefactor \(\alpha\) from our experimental data. In simulations, we sweep $\alpha$ from \qty{1.2e-2} to \qty{2.4e-1} for different irradiation dose for \qty{280}{\nano\meter} pillars, \qty{480}{\nano\meter} pillars, and mesas and plot the results in open triangles in Figure~\ref{figSI_alpha} a,c,e. We arrive at a best fit to our data of  \(\alpha\) = 0.024(2), which we use to plot the simulations in Figure 2 in the main text. 
This small ($<$1) value of $\alpha$ can have many origins 
including nonunity capture probability, Frenkel defect recombination \cite{antonov_statistical_2014,losero_creation_2023} and vacancy cluster formation\cite{kasperczyk_statistically_2020,santonocito_nv_2024}.

\section{Estimation of monovacancy diffusion constant} 
By comparing the number of NV centers formed in 280 nm diameter pillars $N_{NV}^{280}$ and 480 nm diameter pillars $N_{NV}^{480}$ after \(t_{anneal} = \qty{11}{\minute}\), one can estimate the monovacancy diffusion constant, as long as the diffusion length is larger than the pillar size, \textit{i.e.} \(2 \sqrt{2 D_V t_{anneal}} \approx \phi /2\), where \(D_V\) is monovacancy diffusion constant and \(\phi\) is the pillar diameter. One must also consider that compared to bulk diamonds, the number of NVs formed inside nanopillars during vacancy diffusion is suppressed due to vacancy absorption at the pillar sidewalls. 

\begin{figure}
    \centering
    \centerline{\includegraphics[width= 180mm]{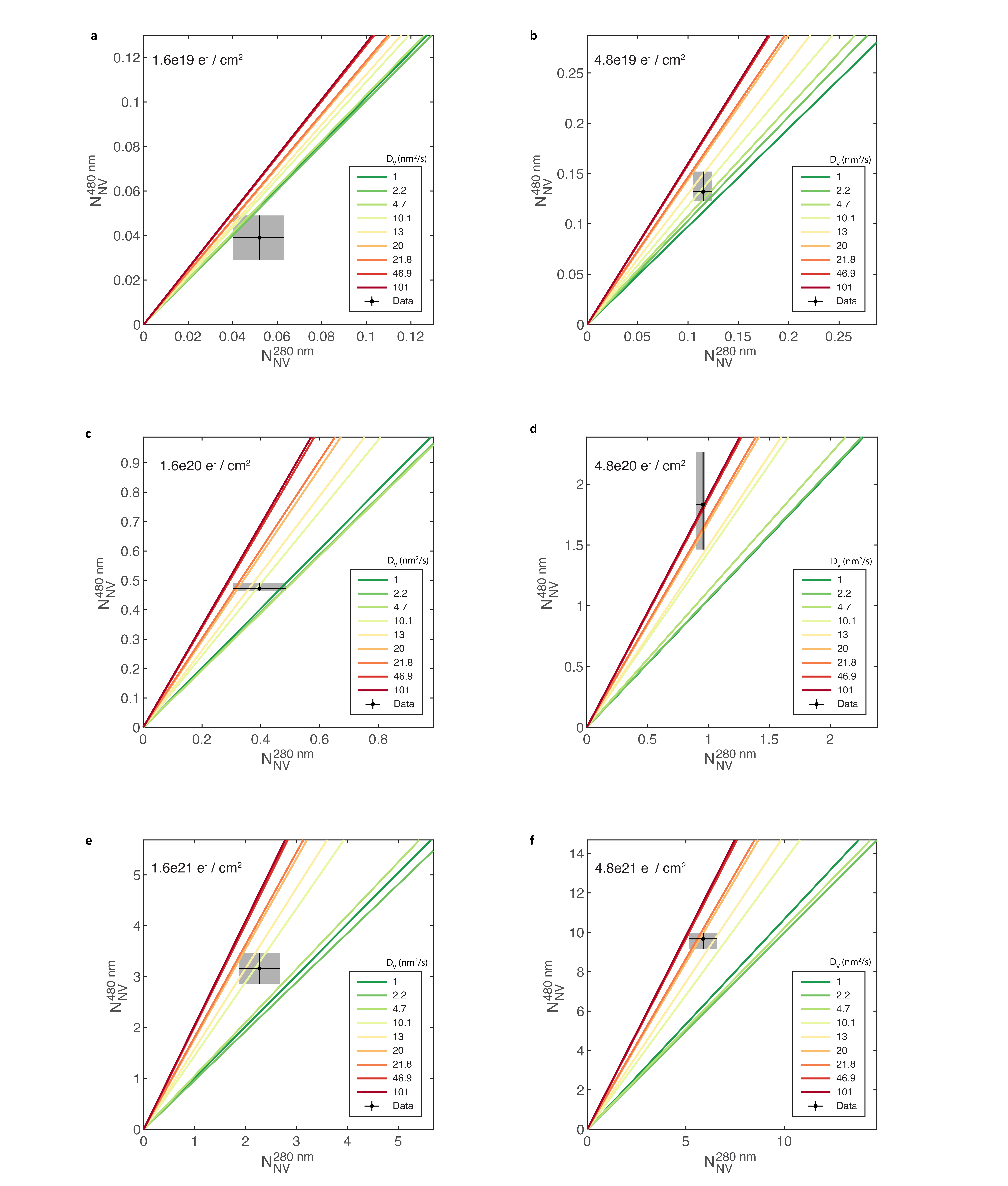}}
    \caption{ Relative number of NVs for two pillar sizes vs. monovacancy diffusion constant. The ratio of the number of NVs for \qty{480}{\nano\meter} pillar to \qty{280}{\nano\meter} pillar $\frac{N_{NV}^{480}}{N_{NV}^{280}}$ is extracted from our MC simulations with varying \(D_V\) for (a) \qty{8}{\pico\coulomb}, (b) \qty{24}{\pico\coulomb}, (c) \qty{80}{\pico\coulomb}, (d) \qty{240}{\pico\coulomb}, (e) \qty{800}{\pico\coulomb}, and (f) \qty{2400}{\pico\coulomb} $\delta$-electron irradiation. Solid lines represent $\frac{N_{NV}^{480}}{N_{NV}^{280}}$ for each \(D_V\). Measured values for \(N_{NV}^{480}\) and \(N_{NV}^{280}\) are plotted with errorbars denoting \qty{95}{\percent} confidence interval. The shaded areas show 2-d \qty{90}{\percent} confidence intervals for a joint probability distribution of a combination \((N_{NV}^{480},N_{NV}^{280})\). }
    \label{figSI_rellambda}
\end{figure}

In Figure~\ref{figSI_rellambda}, we show the relative number of NVs ($\frac{N_{NV}^{480}}{N_{NV}^{280}}$) created for \qty{280}{\nano\meter} and \qty{480}{\nano\meter} diameter pillars (after \(t_{anneal} = \qty{11}{\minute}\)). Overall, the larger pillars host a higher number of created NVs, indicating \(2 \sqrt{2 D_V \qty{660}{\second}} > 280 /2\qty{}{\nano\meter} \). Thus, we characterize the lower bound of \(D_V\) to be \qty{3.7}{\nano\meter\squared\per\second}.

We also use MC simulations to simulate the relative number of NVs for different \(D_V\), as shown in solid lines in Figure~\ref{figSI_rellambda}. 
We find \(\qty{13}{\nano\meter\squared\per\second}<D_V<\qty{20}{\nano\meter\squared\per\second}\) agrees with measured $\frac{N_{NV}^{480}}{N_{NV}^{280}}$ for all irradiation dose.

\section{Characterization of \(\sigma_{sys}\) and \(\sigma_{PSF}\)}
Our method of estimating \(\sigma_{loc}\) requires precise characterization of \(\sigma_{PSF}\) and \(\sigma_{sys}\). To achieve this, we first perform affine transformations on \(40\times\qty{40}{\micro\meter\squared}\) confocal image to account for aberrations in our imaging system. We use a single image of \qty{2400}{\pico\coulomb}-irradiated block to optimize the transformations, where the PL maximum positions of the pillars are recorded. We assume the PL maximum positions are set by the spatial profile of the photonic mode of the pillars and the positions of the pillars are the same as the EBL mask design. Then, the lateral displacement of the EBL mask design from the PL maximum positions can be attributed to optical aberrations of the confocal system, where we minimize the root mean square error (RMSE) of the displacements by using affine transformations on the confocal image. Any residual RMSE after optimizing affine transformations is defined as \(\sigma_{sys}\). We evaluate \(\sigma_{sys}\) by using a different image of \qty{2400}{\pico\coulomb}-irradiated block, which we find to be \qty{41}{\nano\meter}. After characterizing  \(\sigma_{sys}\), all confocal images taken afterward are transformed using the same transformations. 

Then, we characterize  \(\sigma_{PSF}\) by measuring the PSF of a single NV in the mesa region. 6 different NVs are identified in the mesa and imaged using our confocal microscopy. The images are then individually fit using a 2D Gaussian function, from which we evaluate \(\sigma_{PSF} = \qty{235}{\nano\meter}\).

\section{Estimation of monovacancy diffusion constant from \(\sigma_{loc}\)}
We estimate the monovacancy diffusion constant \(D_V\) at \qty{850}{\celsius} by directly comparing \(\sigma_{loc}\) with MC simulation results. In three-dimensional diffusion, the diffusing object can be modeled as a random walker that jumps along the edges of a cubic lattice with a lattice constant \(a\) and a total number of jumps \(N_{jumps}^{tot}\) for a total time \(t_{anneal}\). Then, the diffusion constant is given as \(D = \frac{a^2 N_{jumps}^{tot}}{6t_{anneal}}\). 

\begin{figure}
    \centering
    \centerline{\includegraphics[width= 180mm]{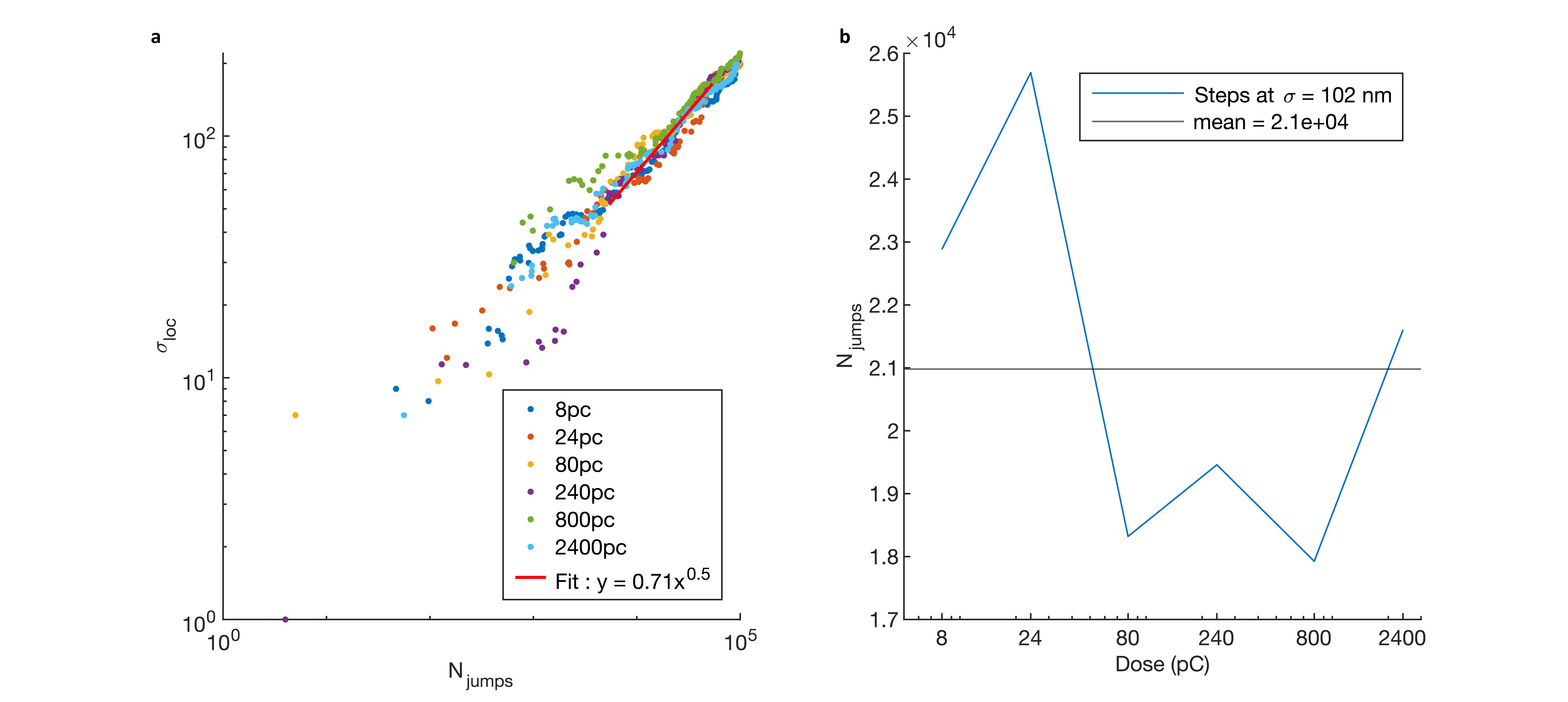}}
    \caption{Extraction of the total number of monovacancy jumps for a given \(\sigma_{loc} \) (a) Calculated \(\sigma_{loc}\) as a function of the number of jumps. At a large number of jumps, \(\sigma_{loc}\) for all doses shows a power law dependence. (b) The number of jumps at \(\sigma_{loc} = \qty{102}{\nano\meter}\) is extracted from (a). The distribution of \(N_{jumps}\) is characterized by a mean of \(2.098\times10^4\).  }
    \label{figSI_sigmaloc}
\end{figure}

In Figure~\ref{figSI_sigmaloc}, we use MC simulations with a simplified model for monovacancy diffusion, where monovacancies perform random walks in a cubic lattice with a lattice constant of \qty{2}{\nano\meter}. We collect NV formation events to extract \(\sigma_{loc}\) for a given number of jumps. Then, we use \(\sigma_{loc}\) characterized from our annealed mesas to estimate \(N_{jumps}^{tot}\) for \(t_{anneal} = \qty{11}{\minute}\). Lastly, we use the above equation to arrive at \(D = \qty{21}{\nano\meter\squared\per\second}\). 

\section{Dipole interaction limit of \(T_2^{Hahn}\)}
 
To understand the source of decoherence in our $\delta$-doped samples, we first consider the contribution of substitutional nitrogen (P1) spins on Hahn echo coherence time \(T_2^{Hahn}\). We consider Ising interactions between P1 spins and the NV center and perform averaging over the ensemble of spin state trajectories and positional randomness\cite{davis_probing_2023}. 
To ensure configurational averaging over many NV center P1 environments (our T2 measurements are all on single NVs), we average multiple single NV Hahn-echo decays together. 
Given the estimated P1 density of \qty{17}{ppm \cdot \nano\meter} from SIMS (the P1 density is typically \(\sim 6\) times less than the nitrogen content determined by SIMS\cite{hughes_two-dimensional_2023}), we calculate the P1-limited \(T_2^{Hahn}\) for an ensemble of NVs to be \(70\sim \qty{90}{\micro\second}\), for a typical correlation times of the P1 spins of $1\sim\qty{2}{\milli\second}$ \cite{hughes_two-dimensional_2023}. 


\section{Saturation photon count measurement}
\begin{figure}
    \centering
    \centerline{\includegraphics[width= 150mm]{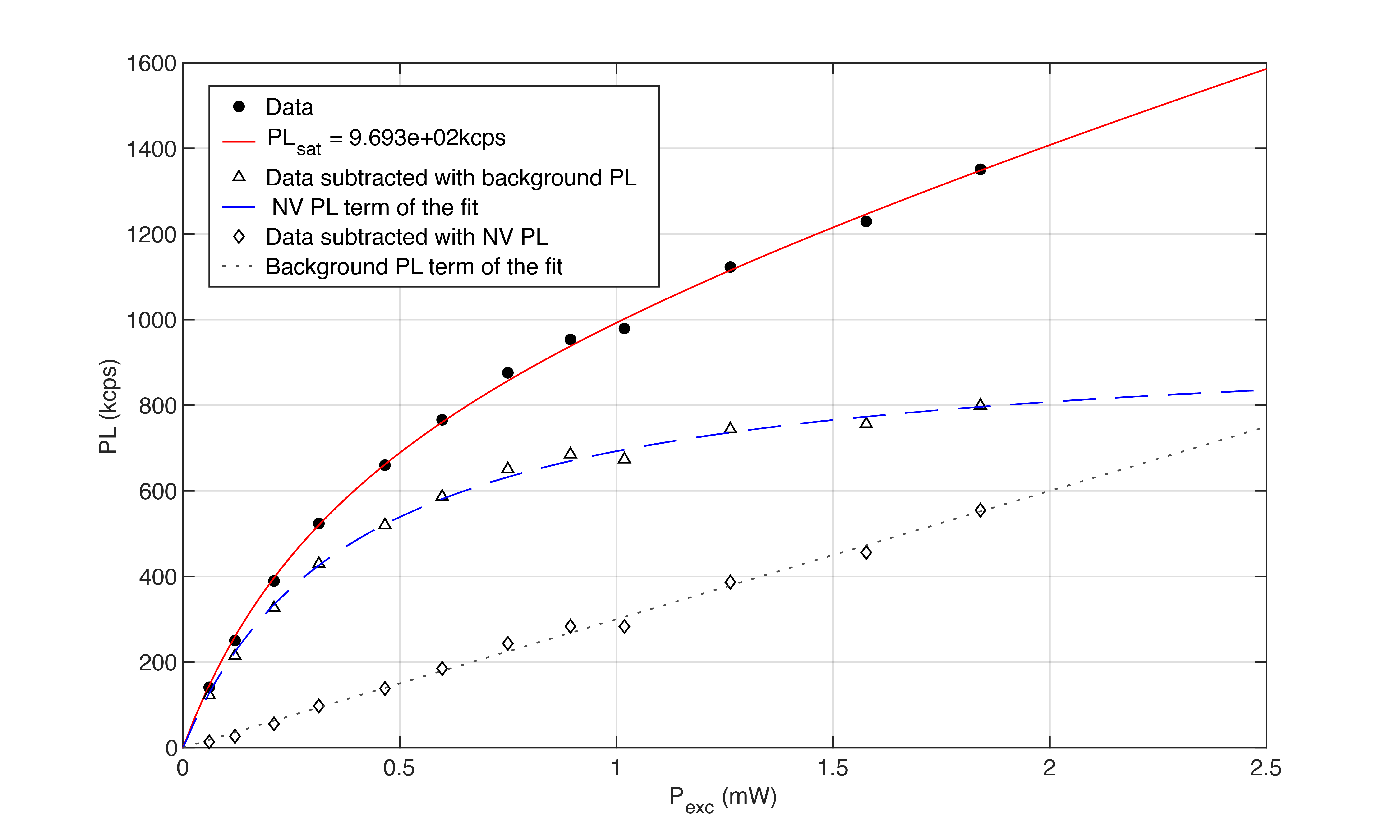}}
    \caption{Photon count rate of a typical single NV pillar as a function of excitation power. The data (black circles) are plotted with the fit curve (red solid line), which is a sum of an NV PL term (blue dashed line) and a background PL term (gray dotted line). We also show the data subtracted with each of the two terms with open triangles and open diamonds. }
    \label{figSI_PLsat}
\end{figure}

We measure \(PL_{sat}\) of nanopillars with single NVs by recording the photon counts \(PL\) as a function of the excitation power \(P_{exc}\). Then, we fit the data with the following:
\begin{equation}
   PL = \frac{PL_{sat}}{1+PL_{sat}/\alpha_{NV} P_{exc}}+\alpha_{bg} P_{exc},
\end{equation}
where \(\alpha_{NV}\) and \(\alpha_{bg}\) are linear coefficients. The first term concerns PL from the NV and the second term concerns PL from the background. In Figure~\ref{figSI_PLsat}, we show a typical \(PL_{sat}\) measurement data of a single NV pillar (black circles), from which we extract \(PL_{sat}\) of \qty{969.3}{kcps}. Note that we extract background counts from the fit, which assumes that the power needed to saturate the background is high compared to the NV. However, we do not see additional non-linear behavior besides the NV term up to the highest \(P_{exc}\) that we use, implying that the background is effectively linear. 

\section{Finite-domain time-dependent simulation results}
We perform FDTD simulations to calculate the photon collection efficiency for a given NV lateral displacement. First, we characterize the geometries of the pillars with top diameter \(\phi_{top}\) of \qty{480}{\nano\meter} and \qty{280}{\nano\meter} from scanning electron micrographs. The nanopillars do not have perfectly vertical sidewalls from the dry etch. We measure height \(h\) of \qty{1414}{\micro\meter} with bottom diameter \(\phi_{bottm}\) of \qty{850}{\nano\meter} and \qty{610}{\nano\meter} for \qty{480}{\nano\meter} and \qty{280}{\nano\meter} pillars, respectively. We use these values in the FDTD simulations described in the main text.

\begin{figure}
    \centering
    \centerline{\includegraphics[width= 180mm]{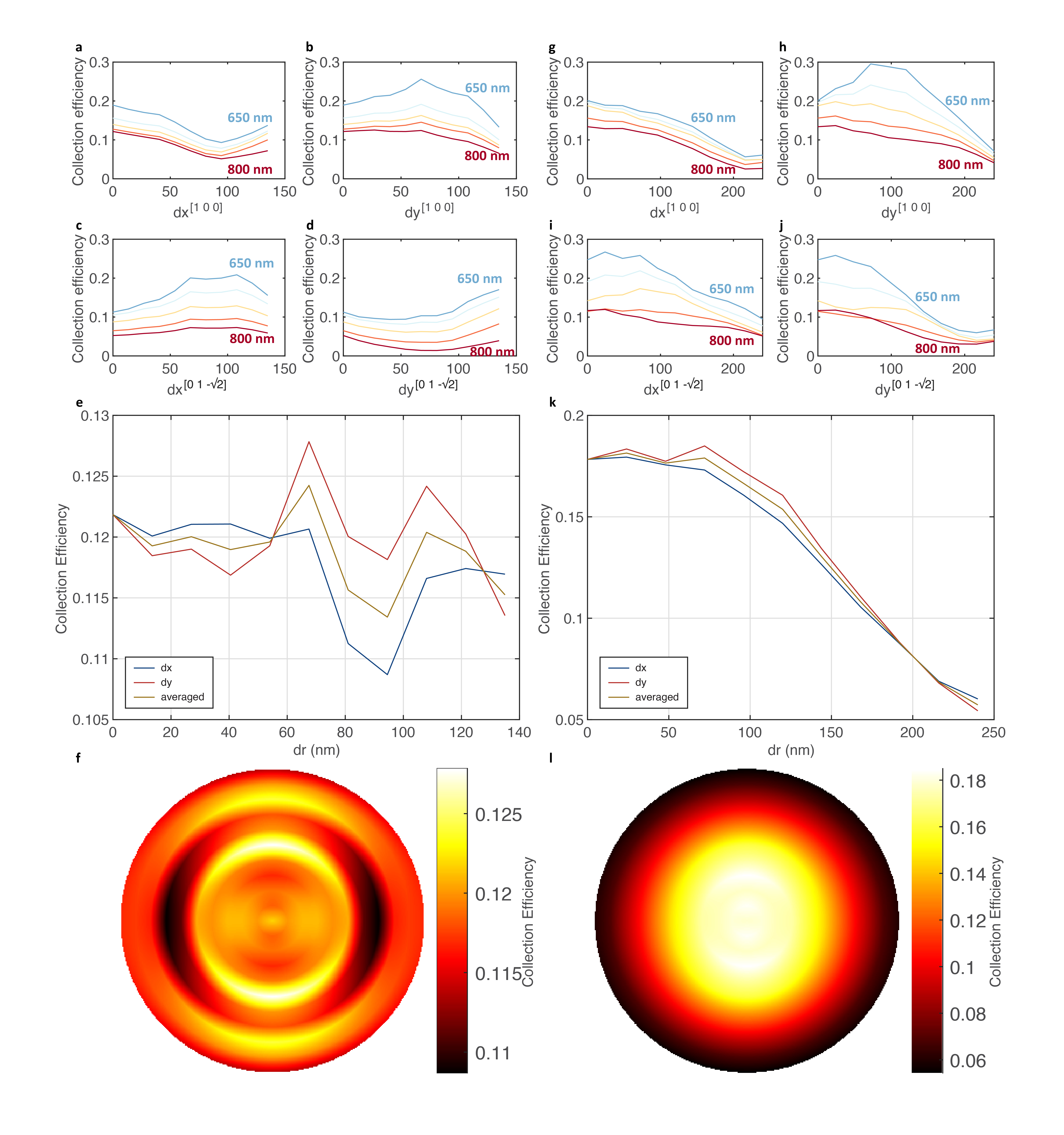}}
    \caption{Collection efficiency of nanopillars as a function of NV lateral displacement. 
    (a-d) The collection efficiency of \qty{280}{\nano\meter} pillar for a given dipole orientation orthogonal to the NV axis is calculated for each emission wavelength for a given orthogonal displacements \(dx\) and \(dy\).
    (e) For each displacement in the \qty{280}{\nano\meter} pillar, we calculate the collection efficiency of an NV by performing a weighted average over the wavelengths and then averaging the two dipole orientations. 
    (f) The 2D map of a collection efficiency inside \qty{280}{\nano\meter} pillar is constructed from (e) as described in the main text.
    (g-j) Same as (a-d) but for \qty{480}{\nano\meter} pillar.
    (k) Same as (e) but for \qty{480}{\nano\meter} pillar.
    (l) Same as (f) but for \qty{480}{\nano\meter} pillar.}
    \label{figSI_FDTD}
\end{figure}

Figure~\ref{figSI_FDTD} shows the calculated collection efficiency of the nanopillars as a function of the NV lateral displacement from the pillar axis. The depth of the dipole is fixed at \qty{53}{\nano\meter} and the NV orientation is set to $[0,\sqrt{2}, 1]$ with respect to the FDTD coordinate system. 
In Figure~\ref{figSI_FDTD} (a-d), we show the collection efficiency of \qty{280}{\nano\meter} pillar as a function of lateral displacement \(dx\) and \(dy\) for a given emission wavelength and dipole orientation. We then average over the wavelength and dipole orientation to calculate collection efficiency for a given NV lateral displacement, as shown in Figure~\ref{figSI_FDTD} (e). 
Figure~\ref{figSI_FDTD} (f) shows the full 2D map of the collection efficiency for an arbitrary lateral displacement inside the \qty{280}{\nano\meter} pillar extrapolated from Figure~\ref{figSI_FDTD} (e), as described in the main text. Likewise, we perform the same simulations for \qty{480}{\nano\meter} pillars, as shown in Figure~\ref{figSI_FDTD} (g-j), to create the same 2D map in Figure~\ref{figSI_FDTD} (l).

We then use the truncated Gaussian function with a spread of \(\sigma_0\) to calculate the mean collection efficiency as a function of \(\sigma_{loc}^{pillar}\) as described in the main text.

\bibliography{references}